\documentclass[12pt]{article}
\pdfoutput=1
\usepackage{amsmath,amssymb,amsthm,amsxtra,overpic,bbm,bm,epsfig,subfigure}
\usepackage{hyperref}
\usepackage{graphicx}
\usepackage{color}
\usepackage{array}
\usepackage{comment}
\usepackage{epstopdf}
\usepackage{float}
\usepackage{multirow}
\numberwithin{equation}{section}
\usepackage{cite}
\usepackage{hyperref}
\usepackage{slashed,stmaryrd}
\usepackage{bbm}
\usepackage{lscape}%
\usepackage{array}
\usepackage{booktabs}%
\usepackage{enumerate}
\usepackage{makecell}
\allowdisplaybreaks[3]

\def\thefootnote{\fnsymbol{footnote}}

\begin{document}
	
\vspace{0.2cm}
	
\begin{center}
{\Large\bf Accidental symmetries in the scalar potential of the Standard Model extended with two Higgs triplets}
\end{center}
	
\vspace{0.2cm}
	
\begin{center}
{\bf Xin Wang}~$^{a,~b}$~\footnote{E-mail: wangx@ihep.ac.cn},
\quad
{\bf Yilin Wang}~$^{a,~b}$~\footnote{E-mail: wangyilin@ihep.ac.cn},
\quad
{\bf Shun Zhou}~$^{a,~b}$~\footnote{E-mail: zhoush@ihep.ac.cn (corresponding author)}
\\
\vspace{0.2cm}
{\small
$^a$Institute of High Energy Physics, Chinese Academy of Sciences, Beijing 100049, China\\
$^b$School of Physical Sciences, University of Chinese Academy of Sciences, Beijing 100049, China}
\end{center}
	
\vspace{1.5cm}
	
\begin{abstract}
The extension of the Standard Model (SM) with two Higgs triplets offers an appealing way to account for both tiny Majorana neutrino masses via the type-II seesaw mechanism and the cosmological matter-antimatter asymmetry via the triplet leptogenesis. In this paper, we classify all possible accidental symmetries in the scalar potential of the two-Higgs-triplet model (2HTM). Based on the bilinear-field formalism, we show that the maximal symmetry group of the 2HTM potential is ${\rm SO(4)}$ and eight types of accidental symmetries in total can be identified. Furthermore, we examine the impact of the couplings between the SM Higgs doublet and the Higgs triplets on the accidental symmetries. The bounded-from-below conditions on the scalar potential with specific accidental symmetries are also derived. Taking the ${\rm SO(4)}$-invariant scalar potential as an example, we investigate the vacuum structures and the scalar mass spectra of the 2HTM.
\end{abstract}
	
\def\thefootnote{\arabic{footnote}}
\setcounter{footnote}{0}
	
\newpage
	
\section{Introduction}
The discovery of the Higgs boson at the CERN Large Hadron Collider (LHC)~\cite{ATLAS:2012yve,CMS:2012qbp} has proved the validity of the Higgs mechanism for the spontaneous gauge symmetry breaking~\cite{Englert:1964et, Higgs:1964ia, Higgs:1964pj, Guralnik:1964eu, Higgs:1966ev, Kibble:1967sv}, i.e., ${\rm SU}(2)^{}_{\rm L}\otimes {\rm U}(1)^{}_{\rm Y} \to {\rm U}(1)^{}_{\rm em}$, in the Standard Model (SM)~\cite{Glashow:1961tr, Weinberg:1967tq, Salam:1968rm}. Though the SM has been extremely successful in describing the electroweak interactions of elementary particles, it remains to be understood how tiny neutrino masses and the matter-antimatter asymmetry in our Universe are generated in a dynamical way~\cite{Xing:2011zza,Xing:2019vks}. In this connection, one of the most natural extensions of the SM is to introduce two ${\rm SU}(2)^{}_{\rm L}$ Higgs triplets ${\bm \phi}^{}_i \equiv (\xi^1_i, \xi^2_i, \xi^3_i)^{\rm T}$ for $i = 1, 2$ with the same hypercharge $Y = -2$, where $\xi^a_i$ (for $a = 1, 2, 3$) are three components of the Higgs triplet ${\bm \phi}^{}_i$. In such a two-Higgs-triplet model (2HTM), the gauge-invariant Lagrangian relevant for charged-lepton and neutrino masses is given by
\begin{eqnarray}
-{\cal L}^{}_{\rm Y} = \overline{\ell^{}_{\rm L}} Y^{}_l H E^{}_{\rm R} + \frac{1}{2} \overline{\ell^{}_{\rm L}} Y^{}_{\nu 1} {\bm \sigma}\cdot {\bm \phi}^{}_1 {\rm i}\sigma^2 \ell^{\rm c}_{\rm L} + \frac{1}{2} \overline{\ell^{}_{\rm L}} Y^{}_{\nu 2} {\bm \sigma}\cdot {\bm \phi}^{}_2 {\rm i}\sigma^2 \ell^{\rm c}_{\rm L} + {\rm h.c.} \; ,
\label{eq:Laglept}
\end{eqnarray}
where $\ell^{}_{\rm L}$ and $H$ stand respectively for the left-handed lepton doublet and the Higgs doublet, $\ell^{\rm c}_{\rm L} \equiv {\rm i}\gamma^2\gamma^0 \overline{\ell^{}_{\rm L}}^{\rm T}$ denotes the charge conjugate of $\ell^{}_{\rm L}$, and ${\bm \sigma}\equiv (\sigma^1, \sigma^2, \sigma^3)^{\rm T}$ are the Pauli matrices with the transpose ``T" acting only on the three-dimensional representation space.\footnote{For two column vectors ${\bf x} \equiv (x^1, x^2, x^3)^{\rm T}$ and ${\bf y} \equiv (y^1, y^2, y^3)^{\rm T}$ in the three-dimensional representation space of the ${\rm SU}(2)^{}_{\rm L}$ group, one can define the inner product ${\bf x}\cdot {\bf y} \equiv x^i y^i$ and the cross product ${\bf x}\times {\bf y} \equiv \epsilon^{ijk} x^j y^k$, where the summation over the repeated indices is implied and $\epsilon^{ijk}$ denotes the totally antisymmetric Levi-Civita tensor.} After the neutral components of the Higgs triplets  and the SM Higgs doublet acquire their vacuum expectation values (vev's), namely, $\langle {\bm \phi}^{}_i\rangle = \sqrt{2} v^{}_i$ and $\langle H \rangle = v^{}_{\rm H}/\sqrt{2}$ with $\sqrt{v^2_{\rm H} + 4v^2_1 + 4v^2_2} \approx 246~{\rm GeV}$, the gauge symmetry ${\rm SU}(2)^{}_{\rm L}\otimes {\rm U}(1)^{}_{\rm Y}$ is broken down to ${\rm U}(1)^{}_{\rm em}$ and the lepton mass matrices are given by $M^{}_l \equiv Y^{}_l v^{}_{\rm H}/\sqrt{2}$ and $M^{}_\nu = \sqrt{2}Y^{}_{\nu 1} v^{}_1 +\sqrt{2} Y^{}_{\nu 2} v^{}_2$. Therefore, tiny Majorana neutrino masses can be attributed to the small vev's of two Higgs triplets if the magnitudes of the Yukawa couplings $Y^{}_{\nu 1}$ and $Y^{}_{\nu 2}$ are of order one. Since there are two Higgs triplets and one Higgs doublet in the 2HTM, the scalar potential and the vacuum structure are more complicated than those in the SM, as we shall discuss later on.
	
On the other hand, tiny Majorana neutrino masses can also be generated in the canonical type-I seesaw model~\cite{Minkowski:1977sc, Yanagida:1979, GellMan1979, Mohapatra:1979ia}, in which three heavy Majorana neutrinos are introduced into the SM. A salient feature of the type-I seesaw model is to provide an elegant explanation for the matter-antimatter asymmetry in our Universe via thermal leptogenesis~\cite{Fukugita:1986hr}. However, in the type-II seesaw model~\cite{Konetschny:1977bn, Magg:1980ut, Schechter:1980gr, Cheng:1980qt, Lazarides:1980nt, Mohapatra:1980yp} where only one Higgs triplet is added to the SM, it is impossible to guarantee CP violation in the out-of-equilibrium and lepton-number-violating decays of the Higgs triplet in the early Universe. As demonstrated in Refs.~\cite{Ma:1998dx, Hambye:2000ui}, a successful leptogenesis scenario can only be realized by introducing at least two Higgs triplets. This is one of the primary motivations to consider the 2HTM. In addition, with multiple Higgs triplets, the non-trivial spontaneous CP violation may arise from the scalar sector, as shown in Ref.~\cite{Ferreira:2021bdj}.
	
When the scalar sector of the SM is extended with extra Higgs doublets or triplets, the scalar potential leads to much richer phenomenology due to various interaction terms among different Higgs multiplets. If some specific relations among the mass parameters and coupling constants in the scalar potential are fulfilled, accidental symmetries besides the electroweak $\rm{SU(2)}^{}_{\rm L} \otimes {\rm U}(1)^{}_{\rm Y}$ gauge symmetry may emerge and enhance the predictive power of the theory.\footnote{Unlike the baryon or lepton number, which happens to be the symmetry of the classical SM Lagrangian based only on the gauge invariance and renormalizability~\cite{Weinberg:1995mt}, accidental symmetries in the present work actually originate from specific relations among the parameters in the scalar potential.} For instance, the possible accidental symmetries in the scalar potential of the two-Higgs-doublet model (2HDM) have been extensively investigated in the literature~\cite{Ivanov:2006yq, Nishi:2006tg, Ivanov:2005hg, Ginzburg:2004vp, Davidson:2005cw, Ivanov:2007de, Ferreira:2009wh, Ferreira:2010yh, Battye:2011jj, Pilaftsis:2011ed,Grzadkowski:2010dj,Maniatis:2007vn,Ferreira:2010bm,Maniatis:2006fs,Grzadkowski:2016szj,Nishi:2007nh,Maniatis:2009vp,Ferreira:2010hy,Birch-Sykes:2020btk,Bento:2020jei,BhupalDev:2014bir,Darvishi:2019dbh,Darvishi:2020teg,Ferreira:2020ana}. It has been pointed out in Ref.~\cite{Pilaftsis:2011ed} that the maximal symmetry group of the 2HDM potential is ${\rm SO(5)}$, and there are in total thirteen kinds of accidental symmetries that the scalar potential can accommodate, including the Higgs family symmetries~\cite{Ginzburg:2004vp}, the generalized CP symmetries~\cite{Ferreira:2009wh} and their certain combinations.

In this paper, we carry out a systematic study of possible accidental symmetries in the scalar potential of the 2HTM. The main motivation for such a study is two-fold. First, as has already been explained, the 2HTM serves as an interesting framework to generate tiny Majorana neutrino masses via the type-II seesaw mechanism and the cosmological matter-antimatter asymmetry via the triplet leptogenesis. Second, the symmetry analysis of the 2HTM scalar potential should be very different from that in the 2HDM. For the scalar potential of only two Higgs triplets, we have found that the maximal symmetry group is ${\rm SO(4)}$. After a careful analysis of all the subgroups of ${\rm SO(4)}$, we reach the conclusion that there are totally eight types of accidental symmetries in the 2HTM potential. For each type of symmetry, the relations among the scalar mass parameters and coupling constants are explicitly given. Furthermore, we consider the complete scalar potential with both two Higgs triplets and one Higgs doublet and demonstrate how the couplings between them affect the accidental symmetries. The bounded-from-below (BFB) conditions on the potential are derived. Finally, taking the ${\rm SO(4)}$-invariant scalar potential as an example, we explore the minima of the potential and find out the neutral vacuum solutions in an analytical way. The vacuum structure and the scalar mass spectra of the ${\rm SO(4)}$-invariant potential are also discussed.
	
The remaining part of this paper is organized as follows. In Sec.~\ref{sec:max}, we write down the complete form of the scalar potential in the 2HTM and reformulate it in the bilinear-field formalism. A detailed analysis of the accidental symmetries is then presented in Sec.~\ref{sec:cla}. We derive the BFB conditions of the scalar potential in  Sec.~\ref{sec:bfb}, and discuss the neutral vacuum solutions of the 2HTM potential in Sec.~\ref{sec:sol}. Finally we summarize our main conclusions in Sec.~\ref{sec:sum}.

\section{The two-Higgs-triplet model}\label{sec:max}

Let us first briefly recall the basic structure of the 2HTM, in particular the scalar potential, and then the bilinear-field formalism~\cite{Ivanov:2005hg, Nishi:2006tg, Ivanov:2006yq}, in order to establish our notations and conventions. The gauge-invariant Lagrangian for the SM Higgs doublet and two Higgs triplets in the 2HTM reads
\begin{eqnarray}
{\cal L}^{}_{\rm 2HTM} = ({\cal D}^\mu_{} H^{}_{})^\dag_{}({\cal D}^{}_\mu H) + ({\cal D}^\mu_{} {\bm \phi}^{}_{1})^\dag_{} \cdot ({\cal D}^{}_\mu {\bm \phi}^{}_1) + ({\cal D}^\mu_{} {\bm \phi}^{}_{2})^\dag_{} \cdot ({\cal D}^{}_\mu {\bm \phi}^{}_2) - V^{}_{\rm 2HTM}\; ,
\label{eq:lan}
\end{eqnarray}
where the covariant derivative is ${\cal D}^{}_\mu \equiv \partial^{}_{\mu} - {\rm i}g t^i_{} W^i_\mu -{\rm i}g^\prime_{} t^0 B^{}_\mu$ with $g$ and $g^\prime$ being respectively the ${\rm SU}(2)^{}_{\rm L}$ and ${\rm U}(1)^{}_{\rm Y}$ gauge couplings, $t^i$ (for $i = 1, 2, 3$) stand for the proper representation matrices of three ${\rm SU(2)}^{}_{\rm L}$ generators [i.e., the matrix elements are $(t^i)^{}_{jk} = (\sigma^i)^{}_{jk}/2$ with $j, k = 1, 2$ for the Higgs doublet $H$ and $(t^i)^{}_{jk} = -{\rm i}\epsilon^{ijk}$ with $j, k = 1, 2, 3$ for the Higgs triplets ${\bm \phi}^{}_1$ and ${\bm \phi}^{}_2$], and $t^0 = Y/2$ is the ${\rm U}(1)^{}_{\rm Y}$ generator with $Y$ being the hypercharge. Notice that we adopt throughout this paper the adjoint representations for the Higgs triplets, which are more convenient for the study of accidental symmetries in the 2HTM. In addition to the kinetic terms in the Lagrangian in Eq.~(\ref{eq:lan}), the scalar potential $V^{}_{\rm 2HTM}$ can be divided into three parts as below
\begin{eqnarray}
V^{}_{\rm 2HTM} &= & V^{}_{\rm H} + V^{}_{\phi} + V^{}_{{\rm H}\phi} \; . \label{eq:pot}
\end{eqnarray}
The most general forms of the pure-doublet potential $V^{}_{\rm H}$, the pure-triplet potential $V^{}_{\phi}$ and the doublet-triplet-mixing potential $V^{}_{{\rm H}\phi}$ are respectively given by
\begin{eqnarray}
V^{}_{\rm H} &=& -\mu^{2}_{\rm H}H^\dag_{} H + \lambda^{}_{\rm H} (H^\dag_{} H)^2_{} \; , \nonumber \\
V^{}_{\phi} &=& m^2_{11}({\bm \phi}^*_1 \cdot {\bm \phi}^{}_1)  +m^2_{22}({\bm \phi}^*_2 \cdot {\bm \phi}^{}_2) + m^2_{12}({\bm \phi}^*_1 \cdot {\bm \phi}^{}_2) + m^{\ast 2}_{12}({\bm \phi}^{}_1 \cdot {\bm \phi}^*_2)  + \lambda^{}_1  ({\bm \phi}^*_1 \cdot {\bm \phi}^{}_1)^2_{} + \lambda^{}_2  ({\bm \phi}^*_2 \cdot {\bm \phi}^{}_2)^2_{} \nonumber \\
&&  + \lambda^{}_3  ({\bm \phi}^*_1 \cdot {\bm \phi}^{}_1) ({\bm \phi}^*_2 \cdot {\bm \phi}^{}_2) + \lambda^{}_4  ({\bm \phi}^*_1 \cdot {\bm \phi}^{}_2) ({\bm \phi}^{}_1 \cdot {\bm \phi}^*_2)  + \frac{\lambda^{}_5}{2} ({\bm \phi}^*_1 \cdot {\bm \phi}^{}_2)^2_{}  + \frac{\lambda^{\ast}_5}{2} ({\bm \phi}^{}_1 \cdot {\bm \phi}^*_2)^2_{}  \nonumber \\
&& + ({\bm \phi}^*_1 \cdot {\bm \phi}^{}_1) \left[ \lambda^{}_6  ({\bm \phi}^*_1 \cdot {\bm \phi}^{}_2) + \lambda^{\ast}_6 ({\bm \phi}^{}_1 \cdot {\bm \phi}^*_2)\right] + ({\bm \phi}^*_2 \cdot {\bm \phi}^{}_2) \left[ \lambda^{}_7 ({\bm \phi}^*_1 \cdot {\bm \phi}^{}_2) + \lambda^{\ast}_7 ({\bm \phi}^{}_1 \cdot {\bm \phi}^*_2)\right] \nonumber \\
&&  + \lambda^{}_8 ({\bm \phi}^*_1 \cdot {\bm \phi}^{\ast}_1)({\bm \phi}^{}_1 \cdot {\bm \phi}^{}_1) + \lambda^{}_9 ({\bm \phi}^{*}_2 \cdot {\bm \phi}^{\ast}_2)({\bm \phi}^{}_2 \cdot {\bm \phi}^{}_2) + \lambda^{}_{10} ({\bm \phi}^*_1 \cdot {\bm \phi}^\ast_2)({\bm \phi}^{}_1 \cdot {\bm \phi}^{}_2)  \nonumber \\
&&  + \lambda^{}_{11} ({\bm \phi}^*_1 \cdot {\bm \phi}^\ast_1)({\bm \phi}^{}_2 \cdot {\bm \phi}^{}_2) + \lambda^\ast_{11} ({\bm \phi}^{}_1 \cdot {\bm \phi}^{}_1)({\bm \phi}^*_2 \cdot {\bm \phi}^\ast_2) + \lambda^{}_{12} ({\bm \phi}^*_1 \cdot {\bm \phi}^\ast_1)({\bm \phi}^{}_1 \cdot {\bm \phi}^{}_2)  \nonumber \\
&& + \lambda^{\ast}_{12} ({\bm \phi}^{}_1 \cdot {\bm \phi}^{}_1)({\bm \phi}^\ast_1 \cdot {\bm \phi}^*_2) + \lambda^{}_{13} ({\bm \phi}^*_2 \cdot {\bm \phi}^\ast_2)({\bm \phi}^{}_1 \cdot {\bm \phi}^{}_2) + \lambda^{\ast}_{13} ({\bm \phi}^{}_2 \cdot {\bm \phi}^{}_2)({\bm \phi}^*_1  \cdot {\bm \phi}^\ast_2) \; , \nonumber \\
V^{}_{{\rm H}\phi} & = & \lambda^{}_{14}(H^\dag_{}H)({\bm \phi}^*_1 \cdot {\bm \phi}^{}_1) + \lambda^{}_{15}(H^\dag_{}H)({\bm \phi}^*_2 \cdot {\bm \phi}^{}_2) + \lambda^{}_{16}(H^\dag_{}H) ({\bm \phi}^*_1 \cdot {\bm \phi}^{}_2) + \lambda^{\ast}_{16}(H^\dag_{}H) ({\bm \phi}^{}_1 \cdot {\bm \phi}^*_2) \nonumber \\
&& + \lambda^{}_{17}(H^\dag_{}{{\rm i} \bm \sigma} H)\cdot({\bm \phi}^\ast_1 \times {\bm \phi}^{}_1) + \lambda^{}_{18}(H^\dag_{}  {\rm i} {\bm \sigma} H)\cdot({\bm \phi}^\ast_2 \times {\bm \phi}^{}_2) + \lambda^{}_{19}(H^\dag_{}{\rm i} {\bm \sigma} H)\cdot({\bm \phi}^\ast_1 \times {\bm \phi}^{}_2) \nonumber \\
&& +  \lambda^{\ast}_{19}( H^\dag_{} {\rm i} {\bm \sigma} H)\cdot( {\bm \phi}^\ast_2 \times {\bm \phi}^{}_1) + (\mu^{}_1 H^{\rm T}_{}{\rm i}\sigma^{}_2{\bm \sigma}\cdot {\bm \phi}^{}_1 H + \mu^{}_2  H^{\rm T}_{}{\rm i}\sigma^{}_2{\bm \sigma}\cdot {\bm \phi}^{}_2 H + {\rm h.c.} )\; .
\label{eq:potpart}
\end{eqnarray}
As in the type-II seesaw model, the minimization of the full scalar potential gives rise to the vev's of the neutral components of the Higgs triplets, leading to tiny Majorana neutrino masses. To this end, the trilinear coupling terms $\mu^{}_1 H^{\rm T}_{}{\rm i}\sigma^{}_2{\bm \sigma}\cdot {\bm \phi}^{}_{1} H$ and $\mu^{}_2 H^{\rm T}_{}{\rm i}\sigma^{}_2{\bm \sigma}\cdot {\bm \phi}^{}_{2} H$ in the doublet-triplet mixing potential $V^{}_{{\rm H}\phi}$ in Eq.~(\ref{eq:potpart}) are crucially important. Roughly speaking, given the neutrino masses at the sub-eV level, the sizes of the mass parameters $|\mu^{}_1|$ and $|\mu^{}_2|$ must be suppressed in comparison to other mass-dimensional parameters.  Therefore we first switch off the trilinear terms, as well as all the quartic coupling terms in $V^{}_{{\rm H}\phi}$, in order to analyze the accidental symmetries in the pure-triplet potential $V^{}_\phi$. Then, the doublet-triplet-mixing terms will be included and their impact on the accidental symmetries will also be studied.

The pure-triplet potential $V^{}_\phi$ in Eq.~(\ref{eq:potpart}) can be rewritten in the quadratic form by using the bilinear-field formalism~\cite{Ivanov:2005hg, Nishi:2006tg, Ivanov:2006yq}. To be explicit, we introduce the scalar multiplet
\begin{eqnarray}
\Phi \equiv \left(\begin{matrix}
{\bm \phi}^{}_1 \\
{\bm \phi}^{}_2 \\
{\bm \phi}^\ast_1 \\
{\bm \phi}^\ast_2 \\
\end{matrix}\right) \; ,
\label{eq:Phi}
\end{eqnarray}
where all the four components of $\Phi$ transform in the same way under the ${\rm SU(2)}^{}_{\rm L}$ gauge group. This is the direct consequence of the pure imaginary representation matrices $(t^i)_{jk} = -{\rm i}\epsilon^{ijk}$ such that ${\bm \phi}^{(\ast)}_{i} \rightarrow \exp ( -{\rm i} t^j \alpha^{}_j ) {\bm \phi}^{(\ast)}_{i}$, where $\alpha^{}_j$ (for $j = 1,2,3$) are the spacetime-dependent real parameters. Different from the traditional bilinear-field formalism~\cite{Ivanov:2005hg,Nishi:2006tg,Ivanov:2006yq}, the scalar multiplet $\Phi$ constructed in Eq.~(\ref{eq:Phi}) satisfies the  Majorana condition~\cite{Battye:2011jj, Pilaftsis:2011ed}
\begin{eqnarray}
\Phi = {\rm C} \Phi^\ast_{} \; ,
\label{eq:Phi_Maj}	
\end{eqnarray}
where we have defined ${\rm C} \equiv \sigma^{1}_{} \otimes \sigma^{0}_{} \otimes {\bf I}^{}_{3\times3}$ with $\sigma^{0}_{}$ and ${\bf I}^{}_{3\times3}$ being the two- and three-dimensional identity matrices, respectively.\footnote{Since the charge-conjugation operator ${\rm C}$ does not affect the inner space of ${\bm \phi}^{}_{1,2}$ and ${\bm \phi}^\ast_{1,2}$, we simply omit the identity matrix ${\bf I}^{}_{3\times3}$ in the following discussions.} The significant advantage of such a Majorana formalism~\cite{Battye:2011jj, Pilaftsis:2011ed} is that the multiplet $\Phi$ contains both ${\bm \phi}^{}_{1,2}$ and ${\bm \phi}^{\ast}_{1,2}$, which allows us to explore the triplet family symmetries and generalized CP symmetries simultaneously by applying the same transformation to the scalar fields~\cite{Battye:2011jj}. Meanwhile, the four-component form of $\Phi$ is also necessary for us to rewrite the terms, like $({\bm \phi}^*_i \cdot {\bm \phi}^{\ast}_j)({\bm \phi}^{}_k \cdot {\bm \phi}^{}_l)$ for $i, j, k, l = 1, 2$, into the quadratic forms. With the help of the multiplet $\Phi$, one can construct the vector $R^\mu_{}$ in the bilinear-field space as follows
\begin{eqnarray}
R^\mu_{} = \Phi^\dag_{} \Sigma^\mu_{} \Phi \; ,
\label{eq:R_def}
\end{eqnarray}
where $\Sigma^\mu_{} \equiv \Sigma^\mu_{\alpha\beta} \sigma^\alpha_{} \otimes \sigma^\beta_{}$ (for $\alpha,\beta = 0,1,2,3$) and there is no summation over the indices $\alpha$ and $\beta$. If we further require $R^\mu_{}$ to be invariant under the charge conjugation of $\Phi$, then the following constraint on the matrices $\Sigma^{\mu}_{}$ should be fulfilled
\begin{eqnarray}
(\Sigma^\mu_{})^{\rm T}_{} = {\rm C}^{-1}_{} \Sigma^\mu_{} {\rm C} \; .
\label{eq:Sigma}
\end{eqnarray}
Given the definitions of $\Sigma^\mu_{}$ and ${\rm C}$, we can derive from Eq.~(\ref{eq:Sigma}) the identity
\begin{eqnarray}
\Sigma^\mu_{\alpha\beta} = \Sigma^\mu_{\lambda\rho} (\Lambda^{}_1)^\lambda_\alpha (\Lambda^{}_2)^\rho_\beta \; ,
\label{eq:alphabeta}
\end{eqnarray}
with two diagonal matrices $\Lambda^{}_1 \equiv {\rm Diag}\{1,1,1,-1\}$ and $\Lambda^{}_2 \equiv {\rm Diag}\{1,1,-1,1\}$. By identifying the left- and right-hand sides of Eq.~(\ref{eq:alphabeta}), we obtain ten non-zero $\Sigma^\mu_{}$, namely,
\begin{eqnarray}
\Sigma^0_{} &=& +\frac{1}{2} \sigma^0_{} \otimes \sigma^0_{} \; , \quad
\Sigma^1_{} = -\frac{1}{2} \sigma^2_{} \otimes \sigma^3_{} \; , \nonumber \\
\Sigma^2_{} &=& -\frac{1}{2} \sigma^1_{} \otimes \sigma^0_{} \; , \quad
\Sigma^3_{} = +\frac{1}{2} \sigma^2_{} \otimes \sigma^1_{} \; , \nonumber \\ \Sigma^4_{} &=& -\frac{1}{2} \sigma^1_{} \otimes \sigma^3_{} \; , \quad
\Sigma^5_{} = +\frac{1}{2} \sigma^2_{} \otimes \sigma^0_{} \; , \nonumber \\ \Sigma^6_{} &=& +\frac{1}{2} \sigma^1_{} \otimes \sigma^1_{} \; , \quad
\Sigma^7_{} = +\frac{1}{2} \sigma^0_{} \otimes \sigma^1_{} \; , \nonumber \\ \Sigma^8_{} &=& -\frac{1}{2} \sigma^3_{} \otimes \sigma^2_{} \; , \quad
\Sigma^9_{} = +\frac{1}{2} \sigma^0_{} \otimes \sigma^3_{} \; .
\label{eq:Sigmaform}
\end{eqnarray}
Thus the vector $R^\mu_{}$ has ten components (i.e., $\mu = 0, 1, 2, \cdots, 9$) and can be explicitly written as
\begin{eqnarray}
R^\mu_{} = \left(\begin{matrix}
{\bm \phi}^*_1  \cdot {\bm \phi}^{}_1 + {\bm \phi}^*_2 \cdot {\bm \phi}^{}_2 \\
+\dfrac{\rm i}{2}\left({\bm \phi}^*_{1} \cdot {\bm \phi}^\ast_1 - {\bm \phi}^{}_{1} \cdot {\bm \phi}^{}_1 - {\bm \phi}^{*}_2 \cdot {\bm \phi}^{\ast}_2 + {\bm \phi}^{}_2 \cdot {\bm \phi}^{}_2 \right) \\
-\dfrac{\rm 1}{2}\left({\bm \phi}^*_{1} \cdot {\bm \phi}^\ast_1 + {\bm \phi}^{}_{1} \cdot {\bm \phi}^{}_1 + {\bm \phi}^{*}_2 \cdot {\bm \phi}^{\ast}_2 + {\bm \phi}^{}_2 \cdot {\bm \phi}^{}_2 \right) \\
-{\rm i}\left({\bm \phi}^{*}_1 \cdot {\bm \phi}^\ast_2 - {\bm \phi}^{}_1 \cdot {\bm \phi}^{}_2\right) \\
-\dfrac{1}{2}\left({\bm \phi}^*_1 \cdot {\bm \phi}^\ast_1 - {\bm \phi}^*_2 \cdot {\bm \phi}^\ast_2 + {\bm \phi}^{}_1 \cdot {\bm \phi}^{}_1 - {\bm \phi}^{}_2 \cdot {\bm \phi}^{}_2 \right) \\
-\dfrac{\rm i}{2}\left({\bm \phi}^*_1 \cdot {\bm \phi}^\ast_1 + {\bm \phi}^*_2 \cdot {\bm \phi}^\ast_2 - {\bm \phi}^{}_1 \cdot {\bm \phi}^{}_1 - {\bm \phi}^{}_2 \cdot {\bm \phi}^{}_2 \right) \\
{\bm \phi}^{*}_1 \cdot {\bm \phi}^\ast_2 + {\bm \phi}^{}_1 \cdot {\bm \phi}^{}_2 \\
{\bm \phi}^*_1 \cdot {\bm \phi}^{}_2 + {\bm \phi}^{}_1 \cdot {\bm \phi}^*_2 \\
{\rm i}\left({\bm \phi}^*_1 \cdot {\bm \phi}^{}_2 - {\bm \phi}^{}_1 \cdot {\bm \phi}^*_2 \right) \\
{\bm \phi}^*_1 \cdot {\bm \phi}^{}_1 - {\bm \phi}^*_2 \cdot {\bm \phi}^{}_2 \\
\end{matrix}\right)  \; .
\label{eq:R_form}
\end{eqnarray}
In terms of $R^\mu_{}$, we can recast the pure-triplet potential $V^{}_{\phi}$ into the quadratic form
\begin{eqnarray}
V^{}_{\phi} = \frac{1}{2} M^{}_\mu R^\mu + \frac{1}{4}  L^{}_{\mu\nu} R^\mu_{} R^\nu_{} \; ,
\label{eq:pot2}
\end{eqnarray}
where the coefficient matrices $M^{}_\mu$ and $L^{}_{\mu\nu}$ are respectively given by
\begin{eqnarray}
M^{}_\mu = \left(\begin{matrix}
		m^2_{11}+m^2_{22}&&~ 0&&~ 0&&~ 0&&~ 0&&~ 0&&~ 0 &&~ 2\,{\rm Re}\,m^2_{12} &&~ 2\,{\rm Im}\,m^2_{12} &&~ m^2_{11}-m^2_{22}
\end{matrix}\right) \; ,
\end{eqnarray}
	and
\begin{eqnarray}
L^{}_{\mu \nu} = \left(
\begin{matrix}
			\lambda^{}_1 + \lambda^{}_2 + \lambda^{}_3 && {\bf 0}^{}_{1\times 3} && {\bf 0}^{}_{1\times 3} && P \\
			{\bf 0}^{}_{3\times 1} && K^{}_1 && K^{}_2 && {\bf 0}^{}_{3\times 3}  \\
			{\bf 0}^{}_{3\times 1} && K^{\rm T}_2 && K^{}_1 && {\bf 0}^{}_{3\times 3}  \\
			P^{\rm T}_{} && {\bf 0}^{}_{3\times 3} && {\bf 0}^{}_{3\times 3} && Q \\
\end{matrix}\right) \; ,
\label{eq:Ldef}
\end{eqnarray}
with the relevant matrices defined as
\begin{eqnarray}
P &\equiv& \left(\begin{matrix}{\rm Re}(\lambda^{}_6 +\lambda^{}_7) &&& {\rm Im}(\lambda^{}_6 +\lambda^{}_7) &&& \lambda^{}_1-\lambda^{}_2 \end{matrix} \right) \; , \nonumber \\
K^{}_1 &\equiv& \left(
\begin{matrix}
\lambda^{}_8+\lambda^{}_9-2\, {\rm Re}\,\lambda^{}_{11} & -2\, {\rm Im}\,\lambda^{}_{11} & -{\rm Re}(\lambda^{}_{12}-\lambda^{}_{13}) \\
-2\, {\rm Im}\,\lambda^{}_{11} & \lambda^{}_8+\lambda^{}_9 + 2\, {\rm Re}\,\lambda^{}_{11} & -{\rm Im}(\lambda^{}_{12}+\lambda^{}_{13}) \\
-{\rm Re}(\lambda^{}_{12}-\lambda^{}_{13}) & -{\rm Im}(\lambda^{}_{12}+\lambda^{}_{13}) & \lambda^{}_{10} \\
\end{matrix}\right) \; , \nonumber \\
K^{}_2 &\equiv& \left(
\begin{matrix}
0 & -(\lambda^{}_8-\lambda^{}_9) & +{\rm Im}(\lambda^{}_{12}-\lambda^{}_{13}) \\
\lambda^{}_8-\lambda^{}_9 & 0 & -{\rm Re}(\lambda^{}_{12}+\lambda^{}_{13}) \\
-{\rm Im}(\lambda^{}_{12}-\lambda^{}_{13}) & {\rm Re}(\lambda^{}_{12}+\lambda^{}_{13}) & 0 \\
\end{matrix}\right) \; , \nonumber \\
Q &\equiv& \left(
\begin{matrix}
\lambda^{}_4 + {\rm Re}\,\lambda^{}_5 & {\rm Im}\,\lambda^{}_5 & {\rm Re}(\lambda^{}_6-\lambda^{}_7) \\
{\rm Im}\,\lambda^{}_5 & \lambda^{}_4 - {\rm Re}\,\lambda^{}_5 & {\rm Im}(\lambda^{}_6-\lambda^{}_7) \\
{\rm Re}(\lambda^{}_6-\lambda^{}_7) & {\rm Im}(\lambda^{}_6-\lambda^{}_7) & \lambda^{}_1+\lambda^{}_2-\lambda^{}_3\\
\end{matrix}\right) \; .
\label{eq:PKQ}
\end{eqnarray}
Note that the $3\times 3$ matrix $K^{}_2$ itself is antisymmetric (i.e., $K^{\rm T}_2 = - K^{}_2$), while the $10\times 10$ matrix $L^{}_{\mu \nu}$ in Eq.~(\ref{eq:Ldef}) is actually symmetric.

\section{Accidental symmetries}\label{sec:cla}
In this section, we investigate all possible accidental symmetries that the 2HTM potential can accommodate. First, the maximal symmetry group of the 2HTM is identified by taking both the kinetic terms and the potential into consideration. Then, we classify all the accidental symmetries in the scalar potential and establish the relations among the mass parameters and couplings.

\subsection{The maximal symmetry group}\label{subsec:max}
The maximal symmetry group of the 2HTM can be determined by following the group-theory approach as in Ref.~\cite{Pilaftsis:2011ed}. Although we mainly focus on the accidental symmetries of the scalar potential, the symmetry transformations should in the first place keep the kinetic terms invariant. After rewriting the kinetic terms of two Higgs triplets as $(D^{\mu}_{}\Phi)^\dag_{}(D^{}_\mu \Phi)/2$ with $D^\mu_{} \equiv \sigma^0_{} \otimes \sigma^0_{} \otimes {\cal D}^\mu_{}$, one can easily observe that the symmetry transformation $U$ acting on the $\Phi$-space should belong to the unitary group ${\rm U}(4)$, which will be further constrained by the Majorana condition shown in Eq.~(\ref{eq:Phi_Maj}). As a result, the generators $J^a$ (for $a = 1, 2, 3, \cdots, 16$) of the ${\rm U}(4)$ group are subject to the restriction
\begin{eqnarray}
{\rm C}^{-1}_{} J^{a}_{} {\rm C} = - (J_{}^{a})^\ast_{} \; .
\label{eq:gen_cons}
\end{eqnarray}
Given that all the generators of ${\rm U}(4)$ group can be expressed as the direct products of $\sigma^\alpha_{}$ and $\sigma^\beta_{}$ (for $\alpha, \beta = 0, 1, 2, 3$), it is straightforward to verify that there are six generators of the group ${\rm U}(4)$ satisfying the relations in Eq.~(\ref{eq:gen_cons}), i.e.,
\begin{eqnarray}
J_{}^{1} &=& \frac{1}{2}\sigma^3_{} \otimes \sigma^3_{} \; , \quad J^{2}_{} = \frac{1}{2}\sigma^3_{} \otimes \sigma^1_{} \; , \quad J^{3}_{} = \frac{1}{2}\sigma^0_{} \otimes \sigma^2_{} \; , \nonumber \\
J_{}^{4} &=& \frac{1}{2}\sigma^3_{} \otimes \sigma^0_{} \; , \quad J_{}^{5} = \frac{1}{2}\sigma^1_{} \otimes \sigma^2_{} \; , \quad J_{}^{6} = \frac{1}{2}\sigma^2_{} \otimes \sigma^2_{} \; .
\end{eqnarray}
The Lie	algebra for the above six generators turns out to be
\begin{eqnarray}
[J^{i}_{},J^{j}_{}] = {\rm i}\epsilon^{ijk}_{}J^{k}_{} \; , \quad [J^{i+3}_{},J^{j+3}_{}] = {\rm i}\epsilon^{ijk}_{}J^{k+3}_{} \; , \quad [J^{i}_{},J^{j+3}_{}] = 0  \; , \quad ({\rm for}~ i,j,k = 1,2,3)
\label{eq:liealg}
\end{eqnarray}
with $\epsilon^{ijk}_{}$ being the three-dimensional Levi-Civita symbol. Therefore the maximal symmetry group in the $\Phi$-space is isomorphic to ${\rm SU}(2) \otimes {\rm SU}(2)$.

Then we turn to the $R^\mu_{}$ vector and figure out its transformations. Since the zero-component $R^0_{} = {\bm \phi}^*_1 \cdot {\bm \phi}^{}_1 + {\bm \phi}^*_2 \cdot {\bm \phi}^{}_2$ remains unchanged under the unitary transformations on $\Phi$, we can just concentrate on the nine-dimensional ``spatial'' components $R^{i}_{}$. For an infinitesimal ${\rm SU}(2) \otimes {\rm SU}(2)$ transformation on the multiplet $\Phi$, one can verify that $R^i_{}$ will change to $R^i_{} + \delta R^i_{}$ with
\begin{eqnarray}
\delta R^i_{} = {\rm i} \theta^{}_a \Phi^\dag_{} [\Sigma^i, J^a_{}]\Phi = 2\theta^{}_a f^{aij}_{} R^j_{} \; ,
\label{eq:R_infi}
\end{eqnarray}
where $\theta^{}_a$ are arbitrary infinitesimal parameters associated with the generators $J^a_{}$ of the group ${\rm SU}(2) \otimes {\rm SU}(2)$, and $f^{aij}_{}$ are the structure constants of the ${\rm SU}(4)$ group. The representation matrices $T^a_{}$ of all the six generators in the $R^i_{}$-space can be extracted from Eq.~(\ref{eq:R_infi}) as follows
\begin{eqnarray}
(T^a_{})^{}_{ij} = -{\rm i}f^{aij}_{} ={\rm Tr}\left([\Sigma^i_{},J^a_{}] \Sigma^j_{}\right) \; , \quad ({\rm for}~i,j=1,2,\cdots,9)\; .
\label{eq:R-gene}
\end{eqnarray}
With the help of Eq.~(\ref{eq:R-gene}) we can immediately write down the explicit expressions of all the representation matrices of six generators, i.e.,
\begin{eqnarray}
T^{1}_{} &=& \left(\begin{matrix}
			0 & +{\rm i} & 0 & 0 & 0 & 0  & 0 & 0 & 0 \\
			-{\rm i} & 0 & 0 & 0 & 0 & 0  & 0 & 0 & 0 \\
			0 & 0 & 0 & 0 & 0 & 0 &  0 & 0 & 0 \\
			0 & 0 & 0 & 0 & +{\rm i} & 0 &  0 & 0 & 0 \\
			0 & 0 & 0 & -{\rm i} & 0 & 0  & 0 & 0 & 0 \\
			0 & 0 & 0 & 0 & 0 & 0  & 0 & 0 & 0 \\
			0 & 0 & 0 & 0 & 0 & 0  & 0 & +{\rm i} & 0 \\
			0 & 0 & 0 & 0 & 0 & 0  & -{\rm i} & 0 & 0 \\
			0 & 0 & 0 & 0 & 0 & 0 &  0 & 0 & 0 \\
		\end{matrix}\right) \; , \quad
T^{2}_{} = \left(\begin{matrix}
			0 & 0 & 0  & 0 & 0 & 0 & 0 & 0 & 0 \\
			0 & 0 & +{\rm i}  & 0 & 0 & 0 & 0 & 0 & 0 \\
			0 & -{\rm i} & 0  & 0 & 0 & 0 & 0 & 0 & 0 \\
			0 & 0 & 0  & 0 & 0 & 0 & 0 & 0 & 0 \\
			0 & 0 & 0  & 0 & 0 & +{\rm i} & 0 & 0 & 0 \\
			0 & 0 & 0  & 0 & -{\rm i} & 0 & 0 & 0 & 0 \\
			0 & 0 & 0  & 0 & 0 & 0 & 0 & 0 & 0 \\
			0 & 0 & 0  & 0 & 0 & 0 & 0 & 0 & +{\rm i} \\
			0 & 0 & 0  & 0 & 0 & 0 & 0 & -{\rm i} & 0 \\
		\end{matrix}\right) \; , \nonumber \\
T^{3}_{} &=& \left(\begin{matrix}
			0 & 0 & +{\rm i} & 0 & 0 & 0  & 0 & 0 & 0 \\
			0 & 0 & 0 & 0 & 0 & 0 & 0 & 0 & 0 \\
			-{\rm i} & 0 & 0 & 0 & 0 & 0 & 0 & 0 & 0 \\
			0 & 0 & 0 & 0 & 0 & +{\rm i}  & 0 & 0 & 0 \\
			0 & 0 & 0 & 0 & 0 & 0 & 0 & 0 & 0 \\
			0 & 0 & 0 & -{\rm i} & 0 & 0 & 0 & 0 & 0 \\
			0 & 0 & 0 & 0 & 0 & 0 & 0 & 0 & +{\rm i} \\
			0 & 0 & 0 & 0 & 0 & 0 & 0 & 0 & 0 \\
			0 & 0 & 0 & 0 & 0 & 0 & -{\rm i} & 0 & 0 \\
		\end{matrix}\right) \; , \quad
T^{4}_{} = \left(\begin{matrix}
			0 & 0 & 0 & +{\rm i} & 0 & 0 & 0 & 0 & 0  \\
			0 & 0 & 0 & 0 & +{\rm i} & 0 & 0 & 0 & 0  \\
			0 & 0 & 0 & 0 & 0 & +{\rm i} & 0 & 0 & 0  \\
			-{\rm i} & 0 & 0 & 0 & 0 & 0 & 0 & 0 & 0  \\
			0 & -{\rm i} & 0 & 0 & 0 & 0 & 0 & 0 & 0 \\
			0 & 0 & -{\rm i} & 0 & 0 & 0 & 0 & 0 & 0  \\
			0 & 0 & 0 & 0 & 0 & 0 & 0 & 0 & 0 \\
			0 & 0 & 0 & 0 & 0 & 0 & 0 & 0 & 0  \\
			0 & 0 & 0 & 0 & 0 & 0 & 0 & 0 & 0  \\
		\end{matrix}\right) \; , \nonumber \\
T^{5}_{} &=& \left(\begin{matrix}
			0 & 0 & 0 & 0 & 0 & 0 & 0 & 0 & 0 \\
			0 & 0 & 0 & 0 & 0 & 0 & 0 & 0 & 0 \\
			0 & 0 & 0 & 0 & 0 & 0 & 0 & 0 & 0 \\
			0 & 0 & 0 & 0 & 0 & 0 & +{\rm i} & 0 & 0 \\
			0 & 0 & 0 & 0 & 0 & 0 & 0 & +{\rm i} & 0 \\
			0 & 0 & 0 & 0 & 0 & 0 & 0 & 0 & +{\rm i} \\
			0 & 0 & 0 & -{\rm i} & 0 & 0 & 0 & 0 & 0\\
			0 & 0 & 0 & 0 & -{\rm i} & 0 & 0 & 0 & 0 \\
			0 & 0 & 0 & 0 & 0 & -{\rm i} & 0 & 0 & 0 \\
		\end{matrix}\right) \; , \quad
T^{6}_{} = \left(\begin{matrix}
			0 & 0 & 0 & 0 & 0 & 0 &  +{\rm i} & 0 & 0 \\
			0 & 0 & 0 & 0 & 0 & 0 &  0 & +{\rm i} & 0 \\
			0 & 0 & 0 & 0 & 0 & 0 &  0 & 0 & +{\rm i} \\
			0 & 0 & 0 & 0 & 0 & 0 &  0 & 0 & 0 \\
			0 & 0 & 0 & 0 & 0 & 0 &  0 & 0 & 0 \\
			0 & 0 & 0 & 0 & 0 & 0 &  0 & 0 & 0 \\
			-{\rm i} & 0 & 0 & 0 & 0 & 0 & 0 & 0 & 0 \\
			0 & -{\rm i} & 0 & 0 & 0 & 0 & 0 & 0 & 0 \\
			0 & 0 & -{\rm i} & 0 & 0 & 0 & 0 & 0 & 0 \\
\end{matrix}\right) \; .
\end{eqnarray}
It is evident that the spatial components $R^i_{}$ transform as the nine-dimensional representation of the symmetry group ${\rm SO}(4) \simeq [{\rm SU}(2) \times  {\rm SU}(2)]/Z^{}_2$, where the $Z^{}_2$ group arises from the two-to-one correspondence between $\pm \Phi$ and $R^i_{}$. Therefore, the maximal symmetry group ${\rm SO}(4)$ of the 2HTM potential is obviously distinct from that in the 2HDM, which has been identified as ${\rm SO}(5)$ in Refs.~\cite{Battye:2011jj, Pilaftsis:2011ed}.
	
\subsection{Symmetries of the pure-triplet potential}\label{subsec:pure}
Now that the maximal symmetry group of the 2HTM has been determined, all possible accidental symmetries in the scalar potential can be figured out in a systematic way by analyzing the subgroups of ${\rm SO}(4)$. For the pure-triplet potential $V^{}_\phi$, it is more convenient to work in the $R^i_{}$-space, since all the accidental symmetries in the $R^i_{}$-space can be represented by some simple orthogonal transformations. More explicitly, if the pure-triplet potential $V^{}_\phi$ in Eq.~(\ref{eq:pot2}) is invariant under the orthogonal transformation of  $R^i_{}$, i.e., $R^i_{} \rightarrow O^i_{~j}R^j$, the coefficient matrices $M^{}_i$ and $L^{}_{ij}$ should correspondingly satisfy
\begin{eqnarray}
M^{}_i = (O^{\rm T}_{})^{~k}_i M^{}_k \; , \quad L^{}_{ij} =   (O^{\rm T}_{})^{~m}_i (O^{\rm T}_{})^{~n}_j L^{}_{m n} \; ,
\label{eq:ML}
\end{eqnarray}
where the orthogonality condition $O^{i}_{~j} (O^{\rm T})^{~k}_{i} = \delta^k_j$ is implied. Hence the accidental symmetries in $V^{}_{\phi}$ are just the intersection of symmetries in the linear and quadratic terms of $R^i_{}$. Given the representation matrix $O$ of some symmetry, the relations in Eq.~(\ref{eq:ML}) should be satisfied by the coefficients in the scalar potential in Eq.~(\ref{eq:potpart}). Without loss of generality, one can always choose the basis of $R^i$ such that the off-diagonal elements of $L^{}_{ij}$ (for $i,j=1,2,\cdots,9$) in Eq.~(\ref{eq:Ldef}) are all zero, which is equivalent to imposing the following conditions on relevant coefficients
\begin{eqnarray}
{\rm Im}\,\lambda^{}_5 = 0 \; , \quad \lambda^{}_6 = \lambda^{}_7 \; , \quad  \lambda^{}_8 =\lambda^{}_9 \; , \quad  {\rm Im}\,\lambda^{}_{11} = 0 \; , \quad \lambda^{}_{12} = \lambda^{}_{13} = 0 \; .
\label{eq:dia-cond}
\end{eqnarray}
	
For the moment, it is helpful to make a brief comparison with the 2HDM, in which the spatial components of the bilinear vector constructed from the Higgs doublets are simply in the fundamental representation of the maximal symmetry group ${\rm SO(5)}$~\cite{Pilaftsis:2011ed}. Thus the classification of accidental symmetries can be essentially achieved by counting the number of identical eigenvalues of the coefficient matrix $L^{}_{ij}$~\cite{Ivanov:2007de,Pilaftsis:2011ed}. However, the situation is different in the 2HTM case, since the bilinear vector $R^{i}_{}$ in the 2HTM is no longer in the fundamental representation of the group ${\rm SO}(4)$. For this reason, there is no one-to-one correspondence between the components of $R^i_{}$ and the coordinate axes of the four-dimensional Euclidean space.

The six generators of the ${\rm SO}(4)$ group can be divided into two classes, each of which forms the $\mathfrak{so}(3)$ Lie algebra. Therefore we can rearrange the components of $R^i_{}$ into a rank-two tensor $r^{ij}_{}$, whose elements are defined as $r^{ij}_{} \equiv R^{3i+j-3}$ (for $i,j = 1,2,3$), namely,
\begin{eqnarray}
r^{ij}_{} \equiv \left(
\begin{matrix}
	R^1_{} && R^2_{} && R^3_{} \\
	R^4_{} && R^5_{} && R^6_{} \\
	R^7_{} && R^8_{} && R^9_{} \\
\end{matrix}\right) \; .
\label{eq:R-tensor}
\end{eqnarray}
The indices $i$ and $j$ refer to two different ${\rm SO}(3)$ groups generated by $\{J^1_{}, J^2_{}, J^3_{}\}$ and $\{J^4_{}, J^5_{}, J^6_{}\}$, which will be denoted as ${\rm SO}(3)^i_{}$ and ${\rm SO}(3)^j_{}$, respectively. Correspondingly, the spatial components of $M^{}_\mu$ and $L^{}_{\mu\nu}$ can be recast into a rank-two tensor $M^{}_{ij}$ and a rank-four tensor $L^{}_{im,jn}$, i.e.,
\begin{eqnarray}
M^{}_{ij} \equiv \left(
	\begin{matrix}
	0 && 0 && 2\, {\rm Re}\,m^2_{12} \\
	0 && 0 && 2\, {\rm Im}\,m^2_{12} \\
	0 && 0 && m^2_{11}-m^2_{22} \\
	\end{matrix}\right) \; , \quad
L^{}_{im, jn} \equiv \left(
\begin{matrix}
(K^{}_1)^{}_{ij} && (K^{}_2)^{}_{ij} && {\bf 0}^{}_{3\times 3}\\
(K^{\rm T}_2)^{}_{ij} && (K^{}_1)^{}_{ij} && {\bf 0}^{}_{3\times 3} \\
{\bf 0}^{}_{3\times 3} && {\bf 0}^{}_{3\times 3} && Q^{}_{ij} \\
\end{matrix}\right) \; .
\label{eq:ML-tensor}
\end{eqnarray}
Notice that $L^{}_{i 1, j 1} =  L^{}_{i 2, j 2} = (K^{}_1)^{}_{ij}$, $L^{}_{i 1, j 2} = -L^{}_{i 2, j 1} = (K^{}_2)^{}_{ij}$, $L^{}_{i 1, j 3} = L^{}_{i 3, j 1} = L^{}_{i 2, j 3} = L^{}_{i 3, j 2} = 0$, and $L^{}_{i 3, j 3} = Q^{}_{ij}$ should be understood for $L^{}_{im, jn}$ in Eq.~(\ref{eq:ML-tensor}). Then one can immediately recognize that each column (or row) of $r^{ij}_{}$ in Eq.~(\ref{eq:R-tensor}) can be viewed as the fundamental representation of the group ${\rm SO(3)}^i_{}$ [or ${\rm SO(3)}^j_{}$]. Such an identification enables us to implement the similar method in Ref.~\cite{Ivanov:2007de} to find out all the accidental symmetries in the 2HTM.

\subsubsection{${\rm SO(4)}$ symmetry}
	
Let us first concentrate on the continuous symmetries. In the last subsection, we have determined the maximal symmetry group ${\rm SO}(4)$ that can exist in the pure-triplet potential of the 2HTM. Such a symmetry can be realized by requiring $M^{}_{ij} = (K^{}_2)^{}_{ij} = 0$ in Eq.~(\ref{eq:ML-tensor}) and $P = {\bf 0}$ in Eq.~(\ref{eq:PKQ}), and further demanding that the diagonal matrices $K^{}_1$ and $Q$ are identical and both proportional to the identity matrix ${\bf I}^{}_{3\times 3}$. Now we explain how these requirements put restrictive constraints on the mass parameters and coupling constants.

First, in the basis where $L^{}_{ij}$ is diagonal, Eq.~(\ref{eq:dia-cond}) should be satisfied. Then, $M^{}_{ij} = 0$ implies $m^2_{11} = m^2_{22}$ and $m^2_{12} = 0$, while $P = {\bf 0}$ gives rise to $\lambda^{}_1 = \lambda^{}_2$. At this stage, we have
\begin{eqnarray}
K^{}_1 = \left( \begin{matrix} 2\lambda^{}_8 - 2\,{\rm Re}\,\lambda^{}_{11} & 0 & 0 \cr 0 & 2\lambda^{}_8 + 2\,{\rm Re}\,\lambda^{}_{11} & 0 \cr 0 & 0 & \lambda^{}_{10} \end{matrix} \right) \; , ~~~ Q = \left( \begin{matrix} \lambda^{}_4 + {\rm Re}\, \lambda^{}_5 & 0 & 0 \cr 0 & \lambda^{}_4 - {\rm Re}\, \lambda^{}_5 & 0 \cr 0 & 0 & 2\lambda^{}_1 - \lambda^{}_3 \end{matrix} \right)\; . \quad
\label{eq:KQso4}
\end{eqnarray}
Notice that the identities $\lambda^{}_1 = \lambda^{}_2$ and those in Eq.~(\ref{eq:dia-cond}) have been implemented in Eq.~(\ref{eq:KQso4}). Finally, we identify $K^{}_1 = Q = \lambda^{}_4 {\bf I}^{}_{3\times 3}$. In summary, all these requirements lead to the conditions on the mass parameters $m^2_{11} = m^2_{22}$ and $m^2_{12} = 0$, and those on the coupling constants
\begin{eqnarray}
\lambda^{}_1 &=& \lambda^{}_2 \; , \nonumber \\
\lambda^{}_3 &=& 2\lambda^{}_1 - 2\lambda^{}_8 \; , \nonumber \\ \lambda^{}_4 &=& \lambda^{}_{10} = 2\lambda^{}_8 = 2 \lambda^{}_9 \; , \nonumber \\
\lambda^{}_5 &=& \lambda^{}_{6} = \lambda^{}_7 = \lambda^{}_{11} = \lambda^{}_{12} =\lambda^{}_{13} = 0 \; .
\label{eq:so4_cond}
\end{eqnarray}
From Eq.~(\ref{eq:so4_cond}), we can observe that there are only two independent coupling constants that are non vanishing, e.g., $\lambda^{}_1$ and $\lambda^{}_8$. Under the aforementioned conditions, we expect that the pure-triplet potential should be greatly simplified. The reduced form of the potential $V^{}_\phi$ with an ${\rm SO}(4)$ symmetry becomes
\begin{eqnarray}
V^{}_{\phi,\,{\rm SO(4)}} &=& m^2_{11}({\bm \phi}^*_1 \cdot {\bm \phi}^{}_1 + {\bm \phi}^*_2 \cdot {\bm \phi}^{}_2) + \lambda^{}_1  ({\bm \phi}^*_1 \cdot {\bm \phi}^{}_1+{\bm \phi}^*_2 \cdot {\bm \phi}^{}_2)^2_{} \nonumber \\
&~& + 2\lambda^{}_8 \left[ ({\bm \phi}^*_1 \cdot {\bm \phi}^{}_2)({\bm \phi}^{}_1 \cdot {\bm \phi}^*_2)-({\bm \phi}^*_1 \cdot {\bm \phi}^{}_1)({\bm \phi}^*_2 \cdot {\bm \phi}^{}_2) \right] \nonumber \\
&~& + \lambda^{}_8 \left[ ({\bm \phi}^*_1 \cdot {\bm \phi}^*_1)({\bm \phi}^{}_1 \cdot {\bm \phi}^{}_1)+ 2({\bm \phi}^*_1 \cdot {\bm \phi}^*_2)({\bm \phi}^{}_1 \cdot {\bm \phi}^{}_2)+({\bm \phi}^*_2 \cdot {\bm \phi}^*_2)({\bm \phi}^{}_2 \cdot {\bm \phi}^{}_2) \right] \; .
\label{eq:potso4}
\end{eqnarray}
One can also perform the ${\rm SO}(4)$ transformations on the multiplet $\Phi$, namely, $ \Phi \rightarrow \exp ({\rm i} \theta^{}_a J^a_{})\Phi$, and explicitly verify that Eq.~(\ref{eq:potso4}) is indeed invariant under the ${\rm SO}(4)$ transformations.
	
\subsubsection{${\rm O(3)}^i_{} \otimes {\rm O(2)}^j_{}$ symmetry}

Next, the symmetry ${\rm O(3)}^i_{} \otimes {\rm O(2)}^j_{}$ generated by $\{J^1_{}, J^2_{}, J^3_{}, J^{4}_{}\}$ can be obtained if both $K^{}_1$ and $Q$ have three identical eigenvalues, namely, $K^{}_1 = 2\lambda^{}_{8} {\bf I}^{}_{3 \times 3}$ and $Q = \lambda^{}_4 {\bf I}^{}_{3\times 3}$. In this case, arbitrary three-dimensional rotations in the subspace spanned by $\{r^{1j}_{},r^{2j}_{},r^{3j}_{}\}$ (for $j = 1,2,3$) will not affect the scalar potential. Meanwhile, $\{r^{11}_{},r^{21}_{},r^{31}_{}\}$ and $\{r^{12}_{},r^{22}_{},r^{32}_{}\}$ are associated with the same submatrix $K^{}_1$ in $L^{}_{ij}$, allowing an ${\rm O(2)}$ symmetry in the $\{r^{i1}_{},r^{i2}_{}\}$ (for $i = 1,2,3$) space. Therefore, in addition to the conditions $m^2_{11} = m^2_{22}$ and $m^2_{12} = 0$, the ${\rm O}(3)^{i}_{} \otimes {\rm O}(2)^j_{}$-invariant scalar potential further requires
\begin{eqnarray}
\lambda^{}_1 &=& \lambda^{}_2 \; , \nonumber \\
\lambda^{}_3 &=& 2\lambda^{}_1 - \lambda^{}_4 \; , \nonumber \\
\lambda^{}_{10} &=& 2\lambda^{}_8 = 2\lambda^{}_9  \; , \nonumber \\ \lambda^{}_5 &=& \lambda^{}_{6} = \lambda^{}_7 = \lambda^{}_{11} = \lambda^{}_{12} = \lambda^{}_{13} = 0 \; .
\label{eq:o3o2_cond}
\end{eqnarray}
More explicitly, the pure-triplet potential with the ${\rm O}(3)^{i}_{} \otimes {\rm O}(2)^j_{}$ symmetry turns out to be
\begin{eqnarray}
V^{}_{\phi,\,{\rm O(3)}^i_{} \times {\rm {O(2)}}^j_{}} &= &m^2_{11}({\bm \phi}^*_1 \cdot {\bm \phi}^{}_1 + {\bm \phi}^*_2 \cdot {\bm \phi}^{}_2)  +\lambda^{}_1  ({\bm \phi}^*_1 \cdot {\bm \phi}^{}_1+{\bm \phi}^*_2 \cdot {\bm \phi}^{}_2)^2_{} \nonumber \\
&& +2\lambda^{}_4[ ({\bm \phi}^*_1 \cdot {\bm \phi}^{}_2)({\bm \phi}^{}_1 \cdot {\bm \phi}^*_2)-({\bm \phi}^*_1 \cdot {\bm \phi}^{}_1)({\bm \phi}^*_2 \cdot {\bm \phi}^{}_2)] \nonumber \\
&& +\lambda^{}_8 [({\bm \phi}^*_1 \cdot {\bm \phi}^*_1)({\bm \phi}^{}_1 \cdot {\bm \phi}^{}_1)+ 2({\bm \phi}^*_1 \cdot {\bm \phi}^*_2)({\bm \phi}^{}_1 \cdot {\bm \phi}^{}_2)+({\bm \phi}^*_2 \cdot {\bm \phi}^*_2)({\bm \phi}^{}_2 \cdot {\bm \phi}^{}_2)]  \; .
\label{eq:poto3o2}
\end{eqnarray}
In comparison with the ${\rm SO}(4)$-invariant potential in Eq.~(\ref{eq:potso4}), the less symmetric potential in Eq.~(\ref{eq:poto3o2}) involves another independent coupling constant $\lambda^{}_4$.
	
\subsubsection{${\rm O(2)}^i_{} \otimes {\rm O(3)}^j_{}$ and ${\rm O(2)}^i_{} \otimes {\rm O(2)}^j_{}$ symmetries}

In a similar way, one can find that the scalar potential may possess an ${\rm O(2)}^i_{} \otimes {\rm O(3)}^j_{}$ symmetry. In order to maintain the invariance of the scalar potential under the ${\rm O}(3)^{j}_{}$ transformation in the space spanned by $\{r^{i1}_{}, r^{i2}_{}, r^{i3}_{}\}$, one should demand that the first, second and third eigenvalue of $K^{}_1$ is equal to that of $Q$, respectively. Such a requirement leads to $m^2_{11} = m^2_{22}$ and $m^2_{12} = 0$ as before, and also
\begin{eqnarray}
\lambda^{}_1 &=& \lambda^{}_2 \; , \nonumber \\
\lambda^{}_{10} &=& 2\lambda^{}_1 - \lambda^{}_3 \; , \nonumber \\
\lambda^{}_4 &=&  2\lambda^{}_8 = 2\lambda^{}_9  \; , \nonumber \\
{\rm Re}\,\lambda^{}_5 &=& -2\, {\rm Re}\,\lambda^{}_{11} \; , \nonumber \\
{\rm Im}\,\lambda^{}_{5} &=& {\rm Im}\,\lambda^{}_{11} = 0 \; , \nonumber \\
\lambda^{}_{6} &=& \lambda^{}_7 =  \lambda^{}_{12} =\lambda^{}_{13} = 0 \; .
\label{eq:o3_cond}
\end{eqnarray}
On the other hand, the desired ${\rm O(2)}^i_{}$ symmetry in the $\{r^{1i}_{},r^{2i}_{},r^{3i}_{}\}$ space indicates that two eigenvalues of $K^{}_{1}$ are identical, so are those of $Q$. Hence we can derive the following three conditions, each of which can give rise to an ${\rm O(2)}^i_{}$ symmetry.
\begin{itemize}
\item ${\rm Re}\,\lambda^{}_5 = -2\,{\rm Re}\,\lambda^{}_{11} = 0 \; ;$
\item ${\rm Re}\,\lambda^{}_5 = -2\,{\rm Re}\,\lambda^{}_{11} = -2\lambda^{}_1+2\lambda^{}_8+\lambda^{}_3\; ;$
\item ${\rm Re}\,\lambda^{}_5 = -2\,{\rm Re}\,\lambda^{}_{11} = +2\lambda^{}_1-2\lambda^{}_8-\lambda^{}_3\; .$
\end{itemize}
The above conditions bring about the ${\rm O(2)}^i_{}$ symmetries generated by $J^1_{}$, $J^2_{}$ and $J^3_{}$, respectively. Combining Eq.~(\ref{eq:o3_cond}) with any one of those conditions, we obtain the relations among the coefficients in Eq.~(\ref{eq:potpart}) if the scalar potential has the ${\rm O(2)}^i_{} \otimes {\rm O(3)}^j_{}$ symmetry.

If one of the three conditions $\lambda^{}_{10} = 2\lambda^{}_1 - \lambda^{}_3$, $\lambda^{}_4 =  2\lambda^{}_8 = 2\lambda^{}_9$ and ${\rm Re}\,\lambda^{}_5 = -2\,{\rm Re}\,\lambda^{}_{11}$ in Eq.~(\ref{eq:o3_cond}) is not satisfied, the ${\rm O(2)}^i_{} \otimes {\rm O(3)}^j_{}$ symmetry will be reduced to ${\rm O(2)}^i_{} \otimes {\rm O(2)}^j_{}$.
	
\subsubsection{${\rm O(2)}^j_{}$ symmetry}

Finally, it is interesting to mention that there exists a minimal symmetry group ${\rm O(2)}^j_{}$ generated by $J^4_{}$ in the scalar potential, even if no restrictions are imposed on the mass parameters and coupling constants. In fact, the transformations of two Higgs triplets ${\bm \phi}^{}_1$ and ${\bm \phi}^{}_2$ under the generator $J^4_{}$ are found to be ${\bm \phi}^{}_{1,2} \rightarrow \exp(-{\rm i}\theta^{}_{4}/2){\bm \phi}^{}_{1,2}$ with $\theta^{}_4$ being an arbitrary real constant. This is exactly the global version of the ${\rm U(1)^{}_{Y}}$ transformations, under which the scalar potential in Eq.~(\ref{eq:pot}) is automatically invariant. It is obvious that all the continuous symmetries in the previous discussions have already contained the ${\rm U(1)^{}_{Y}}$ symmetry.
	
\subsubsection{$Z^{}_2$ symmetries}

Apart from the continuous symmetries, the scalar potential of the 2HTM also permits discrete $Z^{}_2$ symmetries. Different from the 2HDM, if a $Z^{}_2$ transformation is applied to $\Phi$ in the 2HTM, three elements in a row or a column of $r^{ij}_{}$ will simultaneously flip their signs. More explicitly, we find that there are in total three different patterns of the sign changes for $r^{ij}_{}$ after the $Z^{}_2$ transformation on $\Phi$, i.e.,
\begin{eqnarray}
{\bf (a)}~\left(\begin{matrix}
			+ && + && - \\
			+ && + && - \\
			- && - && + \\
		\end{matrix}\right) \; ,
		\quad
{\bf (b)}~\left(\begin{matrix}
			- && - && - \\
			+ && + && + \\
			- && - && - \\
		\end{matrix}\right) \; ,
		\quad
{\bf (c)}~\left(\begin{matrix}
			+ && - && - \\
			+ && - && - \\
			+ && - && - \\
		\end{matrix}\right) \; .
\label{eq:Rpattern}
\end{eqnarray}
Some comments on the sign changes are in order. Pattern {\bf (a)} in Eq.~(\ref{eq:Rpattern}) represents the scenarios where the signs of the elements in one column and one row of $r^{ij}_{}$ are reversed simultaneously, corresponding to the $Z^{}_2$ transformations like ${\bm \phi}^{}_1 \rightarrow -{\bm \phi}^{}_1$, ${\bm \phi}^{}_2 \rightarrow {\bm \phi}^{}_2$. Pattern {\bf (b)} refers to the case where the elements in two rows of $r^{ij}_{}$ change their signs, which can be realized by transformations such as ${\bm \phi}^{}_1 \rightarrow -{\bm \phi}^{}_2$, ${\bm \phi}^{}_2 \rightarrow {\bm \phi}^{}_1$. For Pattern {\bf (c)}, the signs of two columns of $r^{ij}_{}$ are reversed, which can be induced by the transformations like ${\bm \phi}^{}_1 \rightarrow -{\bm \phi}^*_2$, ${\bm \phi}^{}_2 \rightarrow {\bm \phi}^*_1$.

In general, the accidental symmetries could be the combinations of the $Z^{}_2$ symmetries and the continuous symmetries. To be specific, we define the $Z^{}_2$ transformations in the $R^i_{}$-space as the reflection of a whole row or column of $r^{ij}_{}$. Then we need to investigate whether the continuous symmetries have already covered all the $Z^{}_2$ symmetries existing in the $R^{i}_{}$-space. If not, we should include the $Z^{}_2$ symmetries that have been ignored in the previous analysis. For instance, the ${\rm O(3)}^i_{} \otimes {\rm O(2)}^j_{}$ transformation contains all the three patterns shown in Eq.~(\ref{eq:Rpattern}), so we do not need to supplement the ${\rm O(3)}^i_{} \otimes {\rm O(2)}^j_{}$ with extra $Z^{}_2$ symmetries. However, for the ${\rm O(3)}^j_{}$ symmetry, one can easily verify that it does not contain Pattern {\bf (c)} in Eq.~(\ref{eq:Rpattern}). Thus an additional $Z^{}_2$ symmetry is necessary, indicating that the complete symmetry group should be ${\rm O(3)}^j_{} \otimes Z^{}_2$.
	
\subsection{Symmetries of the full potential}\label{subsec:ent}

In the full scalar potential, there are also doublet-triplet-mixing terms. In this subsection we switch on the coupling terms between $H$ and ${\bm \phi}^{}_{1,2}$, and consider the accidental symmetries of the entire potential. There are three different kinds of terms where the Higgs doublet and triplets are coupled with each other in $V^{}_{{\rm H}\phi}$, including the quartic terms $(H^\dag_{}H)({\bm \phi}^*_{i} \cdot {\bm \phi}^{}_{j} )$ and $(H^\dag_{}{\rm i}{\bm \sigma} H)\cdot({\bm \phi}^\ast_i \times {\bm \phi}^{}_j)$, as well as the trilinear terms $H^{\rm T}_{}{\rm i}\sigma^{}_2{\bm \sigma}\cdot {\bm \phi}^{}_{i} H$. First of all, the terms $(H^\dag_{}H)({\bm \phi}^*_{i} \cdot {\bm \phi}^{}_{j} )$ behave similarly as the bilinear terms ${\bm \phi}^*_{i} \cdot {\bm \phi}^{}_{j} $ under transformations of ${\bm \phi}^{}_i$. As a consequence, we can obtain the relations among $\lambda^{}_{14}$, $\lambda^{}_{15}$ and $\lambda^{}_{16}$ for each accidental symmetry in an analogous way as those among $m^2_{11}$, $m^2_{22}$ and $m^2_{12}$. The maximal symmetry group of $V^{}_{\rm 2HTM}$ after including the $(H^\dag_{}H)({\bm \phi}^*_{i} \cdot {\bm \phi}^{}_{j} )$ terms is still ${\rm SO(4)}$ as long as the conditions $\lambda^{}_{14} = \lambda^{}_{15}$ and $\lambda^{}_{16}=0$ are fulfilled.
	
As for the $(H^\dag_{}{\rm i}{\bm \sigma} H)\cdot({\bm \phi}^\ast_i \times {\bm \phi}^{}_j)$ terms, if $\lambda^{}_{19} \neq 0$ is assumed, the only symmetry that can exist in $V^{}_{\rm 2HTM}$ is the $Z^{}_2$ symmetry. If $\lambda^{}_{17} = \lambda^{}_{18}$ and $\lambda^{}_{19} = 0$ hold, one can observe that $\lambda^{}_{17}(H^\dag_{}{{\rm i} \bm \sigma} H)\cdot({\bm \phi}^\ast_1 \times {\bm \phi}^{}_1 + {\bm \phi}^\ast_2 \times {\bm \phi}^{}_2)$ remains invariant under the transformations induced by the ${\rm SO(4)}$ generators $\{J^{1}_{},J^{2}_{},J^{3}_{},J^{4}_{}\}$. However, this is not the case for the other two generators $J^5_{}$ and $J^6_{}$. For example, under the transformation $\Phi \rightarrow \exp({\rm i}\theta^{}_{5} J^5_{})\Phi$ with $\theta^{}_5$ being an arbitrary real constant, we have
\begin{eqnarray}
		{\bm \phi}^\ast_1 \times {\bm \phi}^{}_1 &\rightarrow& \cos^2_{} \frac{\theta^{}_5}{2} ({\bm \phi}^\ast_1 \times {\bm \phi}^{}_1) + \sin^2_{} \frac{\theta^{}_5}{2} ({\bm \phi}^\ast_2 \times {\bm \phi}^{}_2) + \frac{1}{2}\sin \theta^{}_5({\bm \phi}^\ast_1 \times {\bm \phi}^{\ast}_2 + {\bm \phi}^{}_2 \times {\bm \phi}^{}_1) \; , \nonumber \\
		{\bm \phi}^\ast_2 \times {\bm \phi}^{}_2 &\rightarrow& \sin^2_{} \frac{\theta^{}_5}{2} ({\bm \phi}^\ast_1 \times {\bm \phi}^{}_1) + \cos^2_{} \frac{\theta^{}_5}{2} ({\bm \phi}^\ast_2 \times {\bm \phi}^{}_2) + \frac{1}{2}\sin \theta^{}_5({\bm \phi}^\ast_1 \times {\bm \phi}^{\ast}_2 + {\bm \phi}^{}_2 \times {\bm \phi}^{}_1) \; ,
\label{eq:quarj56}
\end{eqnarray}
leading to ${\bm \phi}^\ast_1 \times {\bm \phi}^{}_1 + {\bm \phi}^\ast_2 \times {\bm \phi}^{}_2 \rightarrow {\bm \phi}^\ast_1 \times {\bm \phi}^{}_1 + {\bm \phi}^\ast_2 \times {\bm \phi}^{}_2 + \sin \theta^{}_5({\bm \phi}^\ast_1 \times {\bm \phi}^{\ast}_2 + {\bm \phi}^{}_2 \times {\bm \phi}^{}_1)$. Thus we conclude that the $(H^\dag_{}{\rm i}{\bm \sigma} H)\cdot({\bm \phi}^\ast_i \times {\bm \phi}^{}_j)$ terms should not be ${\rm SO(4)}$ invariant, and the maximal symmetry group of the scalar potential in this case is actually ${\rm O(3)}^i_{} \otimes O(2)^j_{}$.
	
Finally, let us consider the trilinear terms $H^{\rm T}_{}{\rm i}\sigma^{}_2{\bm \sigma}\cdot {\bm \phi}^{}_{i} H$. Since the accidental symmetries of the scalar potential in the 2HTM exist only in the flavor space of two Higgs triplets, it is evident that all the terms proportional to $H^{\rm T}_{}{\rm i}\sigma^{}_2{\bm \sigma}\cdot {\bm \phi}^{}_{i} H$ should violate all these symmetries except the ${\rm SO(2)}^j_{}$ symmetry, as well as the $Z^{}_2$ symmetry related to  the exchange of ${\bm \phi}^{}_1$ and ${\bm \phi}^{}_2$ when $\mu^{}_1=\mu^{}_2 $. For this reason, we can regard $H^{\rm T}_{}{\rm i}\sigma^{}_2{\bm \sigma}\cdot {\bm \phi}^{}_{i} H $  as small soft symmetry-breaking terms. The smallness of $|\mu^{}_1|$ and $|\mu^{}_2|$ is arguably natural in the sense of the 't Hooft's criterion for naturalness, which states that {\it a quantity in nature should be small only if the underlying theory becomes more symmetric as that quantity tends to zero}~\cite{tHooft:1979rat}. In the absence of the trilinear terms, the symmetry is in fact enhanced. Despite the fact that these soft terms maximally violate the accidental symmetries of the scalar potential, they are phenomenologically important not only in explaining neutrino masses via the type-II seesaw mechanism, but also in avoiding undesired Goldstone bosons after the spontaneous breakdown of continuous accidental symmetries.
	
\subsection{Classification of accidental symmetries}
Thus far we have accomplished the classification of all the accidental symmetries. It turns out that there are in total eight distinct types of accidental symmetries for the 2HTM potential, which together with the corresponding generators and constraints on coupling constants in the scalar potential have been listed in Table~\ref{Table:classification}. It is worthwhile to notice that there are three different generators $J^{1}_{}$, $J^{2}_{}$ and $J^3_{}$ that induce the corresponding ${\rm O(2)}^i_{}$ symmetries. As a consequence, we can find three kinds of relations among the parameters for each type of the ${\rm O(2)}^i_{} \otimes {\rm O(3)}^j_{}$, ${\rm O(2)}^i_{} \otimes {\rm O(2)}^j_{}$ and ${\rm O(2)}^i_{} \otimes {\rm O(2)}^j_{} \otimes Z^{}_2$ symmetries.
	
\section{Bounded-from-below conditions}\label{sec:bfb}

\begin{table}[t!]
	\centering
	\caption{The bounded-from-below (BFB) conditions for the pure-triplet potentials with specific accidental symmetries.}
	\vspace{0.5cm}
	\begin{tabular}{p{100pt}p{350pt}}
		\toprule
		Symmetries & BFB conditions \\
		\midrule
		${\rm SO(4)}$ & $2\lambda_1^{} > \lambda_8^{} > 0$.  \\
		${\rm O(3)}^i_{} \otimes {\rm O(2)}^j_{}$ & $4\lambda_1^{} > \lambda_4^{} > 0$ and $\lambda_8^{} > 0$.  \\
		${\rm O(2)}^i_{} \otimes {\rm O(3)}^j_{}$ & \rule{0.12cm}{0cm}(i)~If ${\rm Re}\,\lambda^{}_5 = -2\,{\rm Re}\,\lambda^{}_{11} = 0$, the BFB conditions will be $2\lambda_1^{} > \pm \lambda_3^{}$, \\
		& \rule{0.54cm}{0cm}~$\lambda_8^{} > 0$; \\
		& (ii)~If ${\rm Re}\,\lambda^{}_5 = -2\,{\rm Re}\,\lambda^{}_{11} = \pm\left(2\lambda^{}_1-2\lambda^{}_8-\lambda^{}_3\right)$, the BFB conditions \\
		& \rule{0.54cm}{0cm}~will be $2\lambda_1^{} + \lambda_3^{} > 0$ and $4\lambda_8^{} > 2\lambda_1^{} - \lambda_3^{} > 0$. \\
		${\rm O(3)}^j_{} \otimes Z^{}_2$ & $2\lambda_1^{} > \pm \lambda_3^{}$ and $\lambda_8^{} > \pm {\rm Re}\,\lambda_{11}^{}$. \\
		${\rm O(2)}^i \otimes O(2)^j_{}$ & For each of the three cases where the potential possesses the ${\rm O(2)}^i \otimes O(2)^j_{}$ symmetry, the BFB conditions are respectively given by \\
		& \rule{0.24cm}{0cm}(i)~$\lambda_1^{} + \lambda_2^{} +\lambda_3^{} > 0$, $\lambda_4^{}, \lambda_8^{}, \lambda_{10}^{} > 0$ and $4\lambda_1^{} \lambda_2^{} - \lambda_3^2 > 0$;\\
		&	\rule{0.12cm}{0cm}(ii)~$2\lambda_1^{} \pm \lambda_3^{} > 0$, $\lambda_8^{} > \pm {\rm Re}\,\lambda_{11}^{}$ and $\left(2\lambda_1^{} +\lambda_3^{}\right)\left(2\lambda_4^{} - 2\lambda_1^{} +\lambda_3^{}\right) >$ \\
		& \rule{0.66cm}{0cm}~$4({\rm Im}\,\lambda_6^{})^2_{}$; \\
		&	(iii)~$2\lambda_1^{} + \lambda_3^{} > 0$, ${\rm Re}\,\lambda_{11}^{} > \pm\lambda_8^{}$ and $\left(2\lambda_1^{} +\lambda_3^{}\right)\left(2\lambda_4^{} - 2\lambda_1^{} +\lambda_3^{}\right) > $ \\
		& \rule{0.66cm}{0cm}~$4({\rm Re}\,\lambda_6^{})^2_{}$. \\
		${\rm O(2)}^i \otimes O(2)^j_{} \otimes Z^{}_2$ & For the three kinds of ${\rm O(2)}^i \otimes O(2)^j_{} \otimes Z^{}_2$-invariant potentials, the BFB conditions are respectively\\
		& \rule{0.24cm}{0cm}(i)~$2\lambda_1^{} \pm \lambda_3^{} > 0$ and $\lambda_4^{}, \lambda_8^{}, \lambda_{10}^{} > 0$;\\
		& \rule{0.12cm}{0cm}(ii)~$2\lambda_1^{} \pm \lambda_3^{} > 0$, $\lambda_8^{} > \pm {\rm Re}\,\lambda_{11}^{}$ and $ 2\lambda_4^{} - 2\lambda_1^{} +\lambda_3^{} > 0$;  \\
		&	(iii)~$2\lambda_1^{} + \lambda_3^{} > 0$, ${\rm Re}\,\lambda_{11}^{} >  \pm \lambda_8^{}$ and $2\lambda_4^{} - 2\lambda_1^{} +\lambda_3^{} > 0$. \\
		${\rm SO(2)}^j_{} \otimes (Z^{}_2)^2_{}$ & $2\lambda_1^{} \pm \lambda_3^{} > 0$, $\lambda_4^{} \pm {\rm Re}\,\lambda_5^{} > 0$, $\lambda_8^{} \pm {\rm Re}\,\lambda_{11}^{} > 0$ and $\lambda_{10}^{} > 0$. \\
		${\rm O(2)}^j_{} \otimes Z^{}_2$ & Combine Eq.~(\ref{eq:bfb_con}) with $m^2_{11}=m^2_{22}$ or $\lambda^{}_1=\lambda^{}_2$ or $m^2_{12} = 0 = \lambda^{}_6 = \lambda^{}_7$.  \\
		\bottomrule
	\end{tabular}
	\label{Table:bfbcon}
\end{table} 

As we have seen, the restrictions on the mass parameters and coupling constants in the scalar potential are necessary to realize accidental symmetries. In fact, for any realistic model, the scalar potential has to be bounded from below (BFB) such that the lowest energy state (or the vacuum) of the system exists. In the previous literature, the BFB conditions of the Higgs potential have been discussed in detail in the type-II seesaw model~\cite{Arhrib:2011uy, Chakrabortty:2013mha, Bonilla:2015eha}, the Georgi-Machacek model~\cite{Hartling:2014zca, Blasi:2017xmc, Moultaka:2020dmb, Azevedo:2020mjg}, and the more general 2HTM~\cite{Moultaka:2020dmb}. Instead of a comprehensive study of BFB conditions in the 2HTM, we aim in this section to figure out how the BFB conditions can be simplified given the existence of accidental symmetries. Similar to the discussions in Sec.~\ref{sec:max} and Sec.~\ref{sec:cla}, we first restrict ourselves to the scalar potential involving only the Higgs triplets, then we investigate the impact of $V^{}_{\rm H}$ and $V^{}_{{\rm H}\phi}$ on the BFB conditions.
	
At large field values, the quartic part of $V^{}_{\rm 2HTM}$ in Eq.~(\ref{eq:pot}) will be dominant, so the conditions for the potential to be bounded from below are equivalent to those for the quartic terms to be strictly positive definite. Since the quartic terms of the pure-triplet potential have been rewritten as the quadratic form in the $R^\mu_{}$-space, the BFB conditions then can be derived by requiring the coefficient matrix $L^{}_{\mu\nu}$ to be positive definite~\cite{Ivanov:2006yq}. As for Hermitian matrices, whether they are positive definite can be judged by utilizing the Sylvester's criterion, which states that an $n \times n$ Hermitian matrix $\Omega$ is positive definite if and only if all the upper-left $i$-by-$i$ corners of $\Omega$ (for $i=1,2,\cdots,n$) have positive determinants. In the basis where the matrix $L^{}_{ij}$ is diagonal, the application of the Sylvester's criterion leads us to the following inequalities
\begin{eqnarray}
		\lambda_{10}^{} &>& 0\;, \nonumber \\
		\lambda_8^{} \pm {\rm Re}\,\lambda_{11}^{} &>& 0\; , \nonumber \\
		\lambda_1^{} + \lambda_2^{} + \lambda_3^{} &>& 0 \; , \nonumber \\
		\left(\lambda_1^{} + \lambda_2^{} + \lambda_3^{}\right)\left(\lambda_4^{} + {\rm Re}\,\lambda_5^{}\right) - 4({\rm Re}\,\lambda_6^{})^2_{} &>& 0 \; , \nonumber \\	
		\left(\lambda_1^{} + \lambda_2^{} + \lambda_3^{}\right)\left[\lambda_4^2 - ({\rm Re}\,\lambda_5^{})^2_{}\right] -4\lambda_4^{} \left|\lambda_6^{}\right|^2_{} +4\, {\rm Re}\,\lambda_5^{} \left[({\rm Re}\,\lambda_6^{})^2_{} - ({\rm Im}\,\lambda_6^{})^2_{}\right] &>& 0 \; , \nonumber \\
		-4\left(\lambda_1^{} + \lambda_2^{} - \lambda_3^{}\right)\left\{\lambda_4^{} \left|\lambda_6^{}\right|^2_{} - {\rm Re}\,\lambda_5^{} \left[({\rm Re}\,\lambda_6^{})^2_{} - ({\rm Im}\,\lambda_6^{})^2_{}\right]\right\} && \nonumber \\
		+ \left(4\lambda_1^{} \lambda_2^{} - \lambda_3^2\right) \left[\lambda_4^2 - ({\rm Re}\,\lambda_5^{})^2_{}\right] &>& 0 \; ,
		\label{eq:bfb_con}
	\end{eqnarray}
which guarantee that the pure-triplet scalar potential is bounded from below. Next we apply the above inequalities to the scalar potentials with specific accidental symmetries. It is straightforward to find the BFB conditions for all the accidental symmetries that we have obtained in Sec.~\ref{sec:cla}. The final results have been summarized in Table~\ref{Table:bfbcon}.
	
It is worth mentioning that the BFB conditions given in Table~\ref{Table:bfbcon} are sufficient but not necessary  since the ten components of $R^\mu_{}$ are not totally independent. Generally speaking, it is complicated to derive the sufficient and necessary BFB conditions in the 2HTM, particularly when the terms involving the Higgs doublet $H$ are also taken into account. However, accidental symmetries of the potential can reduce the number of free parameters to a large extent, and thus allow us to derive simple BFB conditions. Below we take the ${\rm SO(4)}$-invariant potential as an example, and rewrite the quartic part as
\begin{eqnarray}
		V^{(4)}_{\rm SO(4)} = \lambda^{}_{\rm H}(H^\dag_{} H)^2_{} + \left( \lambda^{}_1-\frac{\lambda^{}_8}{2}\right)R^0_{} R^{0}_{} +  \frac{\lambda^{}_8}{2} R^i_{} R^{i}_{}  + \lambda^{}_{14} H^\dag_{} H R^0_{} \; .
\label{eq:SO4pot_quar}
\end{eqnarray}
It is more convenient to introduce two dimensionless parameters $\zeta$ and $\eta$ defined as
\begin{eqnarray}
		\zeta \equiv \frac{R^0_{}}{H^\dag_{} H} \; , \quad \eta \equiv \frac{R^i_{}R^i_{}}{R^0_{}R^0_{}} \; ,
		\label{eq:dimpara}
\end{eqnarray}
where $\zeta \geq 0$ given that both $H^\dag_{} H$ and $R^0_{}$ are non-negative. As a consequence, Eq.~(\ref{eq:SO4pot_quar}) can be recast into
\begin{eqnarray}
		V^{(4)}_{\rm SO(4)} = (H^\dag_{} H)^2 \left[\left(\lambda^{}_1 - \frac{\lambda^{}_8}{2} + \frac{\lambda^{}_8}{2} \eta \right) \zeta^2_{}+ \lambda^{}_{14} \zeta + \lambda^{}_{\rm H} \right] \; .
\label{eq:SO4pot_rew}
\end{eqnarray}
Then the conditions for $V^{(4)}_{\rm SO(4)} $ to be positive definite for $\zeta \geq 0$ can be explicitly found out
	\begin{eqnarray}
		\lambda^{}_{\rm H} &>& 0 \; ; \nonumber \\
		\lambda^{}_1 - \frac{\lambda^{}_8}{2} + \frac{\lambda^{}_8}{2} \eta &>& 0 \; ; \nonumber \\
		\sqrt{\lambda^{}_{\rm H}\left[ 4\lambda^{}_1 + 2(\eta - 1)\lambda^{}_8 \right]}+\lambda^{}_{14} &>& 0 \; .
		\label{eq:SO4pdcond}
	\end{eqnarray}
Note that $\eta$ cannot arbitrarily vary in the range $[0,+\infty)$. After some calculations we actually find $1/3 \leq \eta \leq 3$, so the sufficient and necessary BFB conditions of the ${\rm SO(4)}$-invariant potential are modified to be
\begin{eqnarray}
		\lambda^{}_{\rm H} &>& 0 \; ; \nonumber \\
		\lambda^{}_1 + {\rm min}\left(\lambda^{}_8, -\lambda^{}_8/3 \right) &>& 0 \; ; \nonumber \\
		2\sqrt{\lambda^{}_{\rm H}\left[\lambda^{}_1 + {\rm min}\left(\lambda^{}_8, -\lambda^{}_8/3 \right)\right]}+\lambda^{}_{14} &>& 0 \; .
		\label{eq:BFBcond_mod}
\end{eqnarray}
In a similar way, one can examine the BFB conditions on the scalar potential with other accidental symmetries. For the phenomenological studies of any realistic models, these conditions should be taken into account.

\section{Neutral vacuum solutions}\label{sec:sol}
As an application of the accidental symmetries in the 2HTM, we proceed to derive the analytical expressions of the vacuum solutions for the scalar potential with a maximal ${\rm SO(4)}$ symmetry, and explore the residual symmetries embodied in the vacua after the spontaneous breaking of ${\rm SU(2)^{}_L} \otimes \rm{U(1)^{}_Y}$ gauge symmetry. We first consider the scalar potential without the soft symmetry-breaking terms, which is also phenomenologically interesting since it provides us with suitable dark matter candidates~\cite{Cirelli:2007xd, FileviezPerez:2008bj, Hambye:2009pw, Chiang:2020rcv}. The complete ${\rm SO(4)}$-invariant scalar potential reads
\begin{eqnarray}
		V^{}_{\rm SO(4)} &= &-\mu^2_{\rm H}H^\dag_{}H + \lambda^{}_{\rm H}(H^\dag_{}H)^2_{} + m^2_{11}({\bm \phi}^*_1 \cdot {\bm \phi}^{}_1 + {\bm \phi}^*_2 \cdot {\bm \phi}^{}_2) + \lambda^{}_{14}H^\dag_{}H({\bm \phi}^*_1 \cdot {\bm \phi}^{}_1 + {\bm \phi}^*_2 \cdot {\bm \phi}^{}_2) \nonumber \\
&& +\lambda^{}_1  ({\bm \phi}^*_1 \cdot {\bm \phi}^{}_1+{\bm \phi}^*_2 \cdot {\bm \phi}^{}_2)^2_{} + 2 \lambda^{}_8 [({\bm \phi}^*_1 \cdot {\bm \phi}^{}_2) ({\bm \phi}^{}_1 \cdot {\bm \phi}^*_2)-({\bm \phi}^*_1 \cdot {\bm \phi}^{}_1) ({\bm \phi}^*_2 \cdot {\bm \phi}^{}_2)] \nonumber \\
		&&+\lambda^{}_8 [({\bm \phi}^*_1 \cdot {\bm \phi}^*_1)({\bm \phi}^{}_1 \cdot {\bm \phi}^{}_1)+2({\bm \phi}^*_1 \cdot {\bm \phi}^*_2)({\bm \phi}^{}_1 \cdot {\bm \phi}^{}_2)+ ({\bm \phi}^*_2 \cdot {\bm \phi}^*_2)({\bm \phi}^{}_2 \cdot {\bm \phi}^{}_2)] \; .
\label{eq:potso4com}	
\end{eqnarray}
	
Since the charge-breaking vacua violate the $U(1)^{}_{\rm em}$ gauge symmetry, we shall focus only on the neutral vacuum solutions, namely, the vev's of $H$, ${\bm \phi}^{}_1$ and ${\bm \phi}^{}_2$ are respectively assigned to be
\begin{eqnarray}
		\langle H \rangle = \frac{1}{\sqrt{2}}
		\begin{pmatrix}
			0 \\ v^{}_{\rm H}
		\end{pmatrix} \; , \quad
		\langle {\bm \phi}^{}_{1} \rangle = \frac{e^{{\rm i}\alpha^{}_1}_{}}{\sqrt{2}}
		\begin{pmatrix}
			v_1^{} \\ {\rm i} v_1^{} \\ 0
		\end{pmatrix}\; ,
		\quad \langle {\bm \phi}^{}_{2} \rangle = \frac{e^{{\rm i}\alpha^{}_2}_{}}{\sqrt{2}}
		\begin{pmatrix}
			v_2^{} \\ {\rm i} v_2^{} \\ 0
		\end{pmatrix} \; ,
		\label{eq:phivev}
\end{eqnarray}
with the parameters $v^{}_{\rm H}$, $v^{}_1$, $v^{}_2$, $\alpha^{}_1$ and $\alpha^{}_2$ all being real. Notice that the vev's $v^{}_{\rm H}$, $v^{}_{1}$ and $v^{}_{2}$ should satisfy $\sqrt{v^2_{\rm H} + 4 v^2_1 + 4 v^2_2}$ $\simeq 246~{\rm GeV}$ in order to realize the spontaneous breaking of the electroweak symmetry at the true energy scale. In addition, $v^{}_1$ and $v^{}_2$ should not be too large (i.e., typically no more than several ${\rm GeV}$) to contradict the experimental upper bound from the measurement of the $\rho$ parameter~\cite{PDG2020}. Substituting these vev's into Eq.~(\ref{eq:potso4com}), we can evaluate the minimum of the potential
\begin{eqnarray}
		V^{}_{\rm SO(4)}(v^{}_{\rm H},v^{}_1,v^{}_2) &=& -\frac{\mu^2_{\rm H}}{2} v^2_{\rm H} +m^2_{11} \left(v^2_{1} + v^2_{2}\right) + \frac{\lambda^{}_{\rm H}}{4}v^4_{\rm H}+\lambda^{}_1  \left(v^2_{1} + v^2_{2}\right)^2_{}+\frac{\lambda^{}_{14}}{2} v^2_{\rm H} \left(v^2_{1} + v^2_{2}\right)  \; , \quad
\label{eq:potso4vev}
\end{eqnarray}
by requiring
\begin{eqnarray}
		0 &=& \frac{\partial}{\partial  v{}_{\rm H}}V^{}_{\rm SO(4)}(v^{}_{\rm H},v^{}_1,v^{}_2)= -\mu^{2}_{\rm H}v^{}_{\rm H} + \lambda^{}_{\rm H}v^3_{\rm H} + \lambda^{}_{14} \left(v^2_{1} + v^2_{2}\right) v^{}_{\rm H} \; , \nonumber \\
		0 &=& \frac{\partial}{\partial  v{}_{1}}V^{}_{\rm SO(4)}(v^{}_{\rm H},v^{}_1,v^{}_2) =  2 m^2_{11} v^{}_1 +4\lambda^{}_1 \left(v^2_{1} + v^2_{2}\right) v^{}_1 + \lambda^{}_{14}v^2_{\rm H}v^{}_1 \; , \nonumber \\
		0 &=& \frac{\partial}{\partial  v{}_{2}}V^{}_{\rm SO(4)}(v^{}_{\rm H},v^{}_1,v^{}_2) =  2 m^2_{11} v^{}_2 +4\lambda^{}_1 \left(v^2_{1} + v^2_{2}\right) v^{}_2 + \lambda^{}_{14}v^2_{\rm H}v^{}_2 \; .
\label{eq:so4vevcond}
\end{eqnarray}

\begin{figure}[t!]
	\centering		\includegraphics[width=\textwidth]{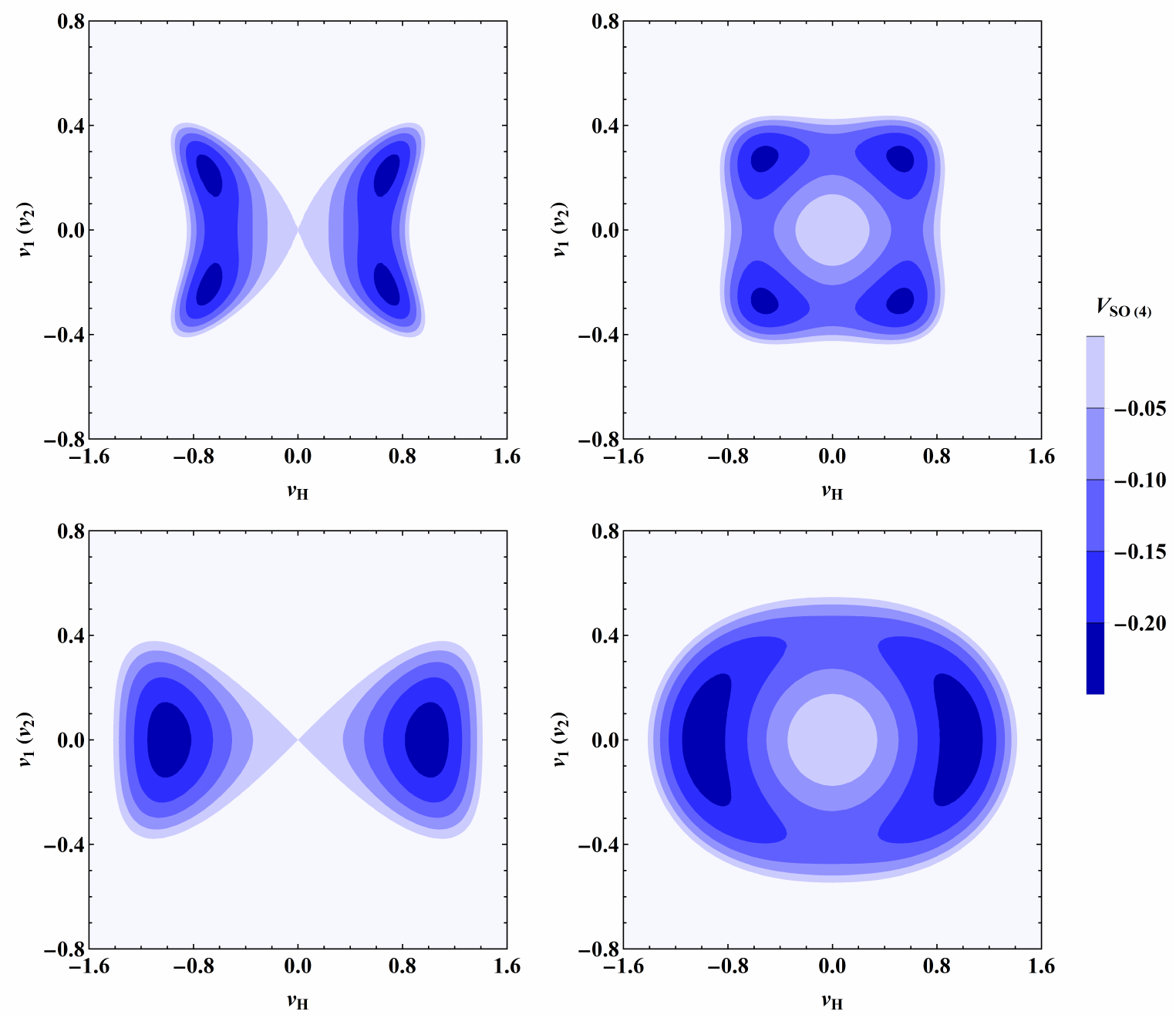} 	
	\caption{The distributions of $V^{}_{\rm SO(4)}$ with respect to the vacuum expectation values $v^{}_{\rm H}$, $v^{}_1$ and $v^{}_2$ in four different cases with some typical parameters. {\it Top-left panel}: $\mu^{2}_{\rm H}=1.99$, $m^{2}_{11}=0.33$, $\lambda^{}_{\rm H}=5.48$, $\lambda^{}_{1}=5.48$, $\lambda^{}_{14}=-6.03$; {\it Top-right panel}: $\mu^{2}_{\rm H}=1.41$, $m^{2}_{11}=-1.49$, $\lambda^{}_{\rm H}=4.12$, $\lambda^{}_{1}=4.12$, $\lambda^{}_{14}=1.65$; {\it Bottom-left panel}: $\mu^{2}_{\rm H}=0.90$, $m^{2}_{11}=0.90$, $\lambda^{}_{\rm H}=0.90$, $\lambda^{}_{1}=0.90$, $\lambda^{}_{14}=-0.68$; {\it Bottom-right panel}: $\mu^{2}_{\rm H}=0.90$, $m^{2}_{11}=-0.90$, $\lambda^{}_{\rm H}=0.90$, $\lambda^{}_{1}=1.49$, $\lambda^{}_{14}=2.09$. For illustration, all the mass-dimensional parameters have been properly normalized to be dimensionless. Furthermore, $v^{}_1$ and $v^{}_2$ are assumed to be equal to each other.}
	\label{fig:vso4} 
\end{figure}

It is not difficult to identify that there are three different types of solutions to Eq.~(\ref{eq:so4vevcond}), which will be presented as follows.
\begin{itemize}
\item {\bf Type A}: $v^{}_{\rm H} \neq 0$, $v^{}_{1,2} \neq 0$
\begin{eqnarray}
			v^2_{\rm H} &=& \frac{4\mu^{2}_{\rm H}\lambda^{}_1 + 2m^2_{11} \lambda^{}_{14}}{4\lambda^{}_{\rm H}\lambda^{}_1-\lambda^2_{14}} \; , \nonumber \\
			v^2_{1} &=& \frac{2m^2_{11}\lambda^{}_{\rm H}+\mu^{2}_{\rm H} \lambda^{}_{14}}{\lambda^2_{14}-4\lambda^{}_{\rm H}\lambda^{}_1}\cos^2_{}\theta \; , \nonumber \\
			v^2_{2} &=& \frac{2m^2_{11}\lambda^{}_{\rm H}+\mu^{2}_{\rm H} \lambda^{}_{14}}{\lambda^2_{14}-4\lambda^{}_{\rm H}\lambda^{}_1}\sin^2_{}\theta \; ,
\label{eq:so4vevsol1}
\end{eqnarray}
with the free rotation angle $\theta \in [0,\pi)$. Some comments on Eq.~(\ref{eq:so4vevsol1}) are helpful. First, it is evident that the solution shown in Eq.~(\ref{eq:so4vevsol1}) is valid only for $\lambda^{2}_{14} \neq 4\lambda^{}_{\rm H}\lambda^{}_1$. In the case of $\lambda^{2}_{14} = 4\lambda^{}_{\rm H}\lambda^{}_1$, the solution of $v^{}_{\rm H} \neq 0$ and $v^{}_{1,2} \neq 0$ to Eq.~(\ref{eq:so4vevcond}) is obtainable if and only if $\mu^2_{\rm H}\lambda^{}_{14} = -2m^2_{11}\lambda^{}_{\rm H}$. Actually the solutions in this case satisfy $v^2_{\rm H} + \lambda^{}_{14} (v^{2}_1 + v^{2}_{2})/\lambda^{}_{\rm H} = \mu^2_{\rm H}/\lambda^{}_{\rm H}$ and all the vev's cannot be uniquely identified. Hence this scenario is apparently far from realistic. Second, once the coupling coefficients are given, $v^{}_{\rm H}$ can be uniquely identified, whereas $v^2_{1}$ and $v^2_{2}$ cannot be fully determined but with the summation $v^2_{1} + v^2_{2}$ fixed. This indicates that there is an extra ${\rm SO(2)}$ residual symmetry in the vacuum manifold after the spontaneous breakdown of the ${\rm SU(2)^{}_{L} \otimes U(1)^{}_{Y}}$ gauge symmetry. Third, since the phases  of ${\bm \phi}^{}_1$ and ${\bm \phi}^{}_2$ do not appear in Eq.~(\ref{eq:potso4vev}), $\alpha^{}_1$ and $\alpha^{}_2$ can take arbitrary values. Finally, in order to guarantee that Eq.~(\ref{eq:so4vevsol1}) determines the (global) minimum of the potential, one should examine the second-order derivatives of $V^{}_{\rm SO(4)}$ with respect to $v^{}_{\rm H}$, $v^{}_1$ and $v^{}_2$. Depending on the sign of $m^2_{11}$, there are two separate regions where the formulae of $v^{2}_{\rm H}$, $v^2_1$ and $v^2_2$ given in Eq.~(\ref{eq:so4vevsol1}) are the minimum of the potential, namely, $ - 2\mu^2_{\rm H}\lambda^{}_{1}/m^2_{11}< \lambda^{}_{14} < - 2 m^2_{11}\lambda^{}_{\rm H}/\mu^2_{\rm H}$ with $m^2_{11}>0$, and $\lambda^{}_{14} < {\rm min}\,( -2\mu^2_{\rm H}\lambda^{}_1/m^2_{11}, -2 m^2_{11}\lambda^{}_{\rm H}/\mu^2_{\rm H})$ with $m^2_{11}<0$.  

Substituting Eq.~(\ref{eq:so4vevsol1}) into Eq.~(\ref{eq:potso4vev}), we can get the minimum of $V^{}_{\rm SO(4)}$ as
\begin{eqnarray}
			V^{\rm min}_{\rm SO(4)} = \frac{m^4_{11}\lambda^{}_{\rm H}+\mu^2_{\rm H} m^2_{11}\lambda^{}_{14}+\mu^4_{\rm H}\lambda^{}_1}{\lambda^2_{14}-4\lambda^{}_{\rm H}\lambda^{}_1} \; .
\label{eq:so4min}
\end{eqnarray}
The vacuum configurations of $V^{}_{\rm SO(4)}$ under the parameter settings that lead to {\bf Type A} vacuum solution are presented in the first row of Fig.~\ref{fig:vso4}, where the top-left and top-right panels refer to $m^2_{11} > 0$ and $m^2_{11} < 0$, respectively. As can be seen in these two plots, a salient difference between these two parameter ranges is that $v^{}_{\rm H} = v^{}_1 = v^{}_2 = 0$ corresponds to the local maximum (saddle point) when $m^2_{11} < 0$ ($m^2_{11} > 0$). By adjusting the values of coupling constants, one can in principle achieve correct vev's of the Higgs doublet and triplets, i.e., $v^{}_{\rm H} \simeq 246~{\rm GeV}$ and $v^{}_{1(2)} \lesssim {\cal O}(1)~{\rm GeV}$ from Eq.~(\ref{eq:so4vevsol1}). Nevertheless, this type of solution is unlikely to reflect the realistic vacuum structure, mainly due to the following two facts. On the one hand, as the scalar potential is ${\rm SO(4)}$-invariant, the spontaneous breakdown of the ${\rm SO(4)}$ symmetry will unavoidably induce superfluous Goldstone particles~\cite{Gelmini:1980re}. On the other hand, if we insert Eq.~(\ref{eq:so4vevsol1}) into the second derivative of $V^{}_{\rm SO(4)}$ with respect to $v^{}_1$ (assuming $v^{}_1= v^{}_2$), we will arrive at
\begin{eqnarray}
		\left.\frac{\partial^{2}_{} V^{}_{\rm SO(4)}}{\partial  v^{2}_{1}}\right|^{}_{\rm vac} = \frac{20\lambda^{}_1 (2m^2_{11}\lambda^{}_{\rm H}+\mu^{2}_{\rm H} \lambda^{}_{14})}{\lambda^2_{14}-4\lambda^{}_{\rm H}\lambda^{}_1} \; ,
\label{eq:secder}
\end{eqnarray}
which is proportional to $v^2_1$ in Eq.~(\ref{eq:so4vevsol1}). Then one can immediately realize that the value of $\partial^2_{} V^{}_{\rm SO(4)}/\partial v^2_1$ at the vacuum has to be negligibly small if $v^{}_{1(2)}$ itself is rather small. Eq.~(\ref{eq:so4vevsol1}) also indicates that $V^{}_{\rm SO(4)}$ should be rather flat along the $v^{}_1$ direction near {\bf Type A} vacuum. This point can be well understood via numerical calculations. To be specific, we select proper parameters that yield $v^{}_{\rm H} \simeq 246~{\rm GeV}$ and $v^{}_{1(2)} \lesssim {\cal O}(1)~{\rm GeV}$. Then we fix $v^{}_{\rm H}$ at its vev and analyze the evolution of $V^{}_{\rm SO(4)}$ along the $v^{}_1$ ($v^{}_2$) direction. A benchmark example is exhibited in Fig.~\ref{fig:vactri}, where the value of $V^{}_{\rm SO(4)}$ remains almost constant in the range $-5~{\rm GeV} \lesssim v^{}_{1(2)}  \lesssim 5~{\rm GeV}$. On this account, the vacuum solution in Eq.~(\ref{eq:so4vevsol1}) will be not stable against either explicit or radiative corrections. Therefore, {\bf Type A} vacuum solution should not be the realistic one.

\begin{figure}[t!]
   	\centering		\includegraphics[width=0.45\textwidth]{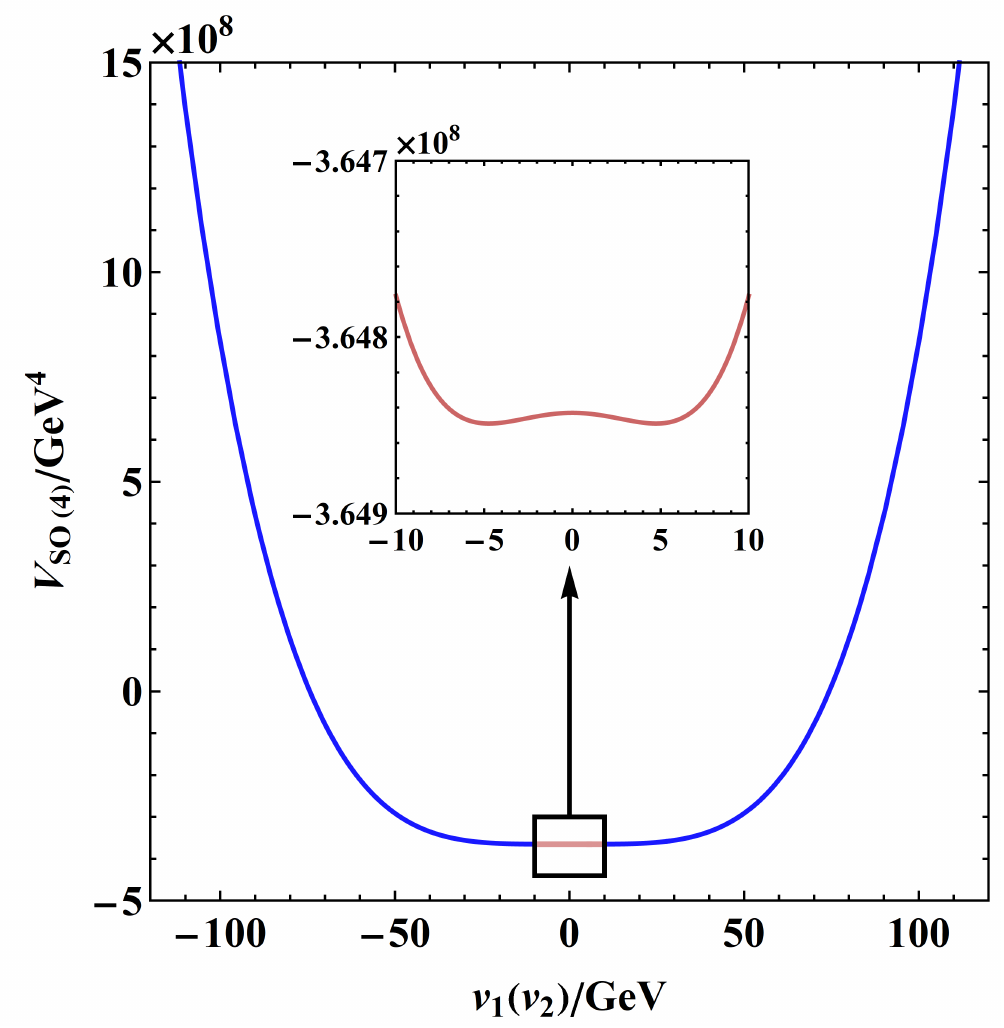} 	\vspace{0cm}
   	\caption{A benchmark example for the evolution of $V^{}_{\rm SO(4)}$ along the $v^{}_{1(2)}$ direction (assuming $v^{}_1 = v^{}_2$), where the coupling coefficients are taken to be $\mu^2_{\rm H} = 2.421 \times 10^4_{}~{\rm GeV^2_{}}$, $m^2_{11}=1.563 \times 10^4_{}~{\rm GeV^2_{}}$, $\lambda^{}_1 = 3.006$, $\lambda^{}_{\rm H} = 0.402$ and $\lambda^{}_{14} = -0.527$, respectively. Notice that $v^{}_{\rm H}$ is fixed at its vev $v^{}_{\rm H} = 245.6~{\rm GeV}$ in the above parameter setting.}
   \label{fig:vactri} 
\end{figure}

\item {\bf Type B}: $v^{}_{\rm H} \neq 0$, $v^{}_{1,2} = 0$
\begin{eqnarray} 
	v^{2}_{\rm H} = \frac{\mu^{2}_{\rm H}}{\lambda^{}_{\rm H}} \; , \quad v^2_1 + v^2_2 = 0 \; ,
\label{eq:so4vevsol2}
\end{eqnarray}
which can be the minimum of the potential if $ - 2m^2_{11}\lambda^{}_{\rm H}/\mu^2_{\rm H} < \lambda^{}_{14} < 0$ with $m^2_{11}>0$ or $\lambda^{}_{14} > -2 m^2_{11}\lambda^{}_{\rm H}/\mu^2_{\rm H}$ with $m^2_{11}<0$. For {\bf Type B} vacuum solution, ${\bm \phi}^{}_{1,2}$ do not acquire non-zero vev's, so the ${\rm SO(4)}$ symmetry can be preserved. The vacuum structures of $V^{}_{\rm SO(4)}$ in this case are presented in the second row of Fig.~\ref{fig:vso4}, where one can observe that there are two degenerate vacua with $\pm |v^{}_{\rm H}|$ regardless of whether $m^{2}_{11} > 0$ or $m^{2}_{11} < 0$. As we shall show later, the modifications from the soft symmetry-breaking terms to Eq.~(\ref{eq:so4vevsol2}) will naturally give rise to tiny but non-vanishing vev's of the Higgs triplets, which satisfy the experimental upper bound from the measurements of the $\rho$ parameter and are suitable for the explanation of neutrino masses via the type-II seesaw mechanism.
\item {\bf Type C}: $v^{}_{\rm H} = 0$, $v^{}_{1,2} \neq 0$
\begin{eqnarray}
      	v^{2}_{\rm H} = 0 \; , \quad v^2_1 + v^2_2 = - \frac{m^2_{11}}{2\lambda^{}_1} \; ,
\label{eq:so4vevsol3}
\end{eqnarray}
        which could be the vacuum solution only if $m^2_{11}<0$ and $\lambda^{}_{14} > - 2\mu^2_{\rm H}\lambda^{}_1/m^2_{11}$. However, since the Higgs doublet does not gain the non-zero vev, Eq.~(\ref{eq:so4vevsol3}) should not be the physically true solution. Therefore we will not consider this vacuum solution in the following discussions.
\end{itemize}

Once one of the above three types of vacua is taken, the number of independent coupling constants in the scalar potential will be further reduced. The complete ${\rm SO(4)}$-invariant potential in Eq.~(\ref{eq:potso4com}) contains six free parameters. As for the {\bf Type A} vacuum, nonzero $v^2_{\rm H}$ and $v^2_1+v^2_2$ impose two constraints. As a result, there will be four free parameters remaining in the scalar potential. While for {\bf Type B} and {\bf Type C} vacua, either $v^2_{1} + v^2_{2}$ or $v^2_{\rm H}$ will be vanishing, leaving five independent coefficients in $V^{}_{\rm SO(4)}$.  Meanwhile, since the number of free parameters in the potential is quite few, the physical scalar bosons with degenerate or vanishing masses are expected to exist. However, if radiative corrections are taken into account, the degeneracy in the scalar masses may be broken and nonzero scalar masses could be generated.

Then we add the soft symmetry-breaking terms $H^{\rm T}_{}{\rm i}\sigma^{}_2{\bm \sigma}\cdot {\bm \phi}^{}_{i} H$ into Eq.~(\ref{eq:potso4com}), which will bring in important corrections to the above vacuum solutions. For clarity, we investigate the following scalar potential
\begin{eqnarray}
V^{}_{\rm soft} = V^{}_{\rm SO(4)} + \mu^{}_1 [H^{\rm T}_{}{\rm i}\sigma^{}_2{\bm \sigma}\cdot ({\bm \phi}^{}_{1} + {\bm \phi}^{}_{2}) H + {\rm h.c.}] \; ,
\label{eq:potsoft}
\end{eqnarray}
where $V^{}_{\rm SO(4)}$ has been given in Eq.~(\ref{eq:potso4com}) and $\mu^{}_1  = \mu^{}_2$ has been assumed. It is evident that the entire potential still possesses a $Z^{}_{2}$ symmetry. Similarly, we assign the vev's of the Higgs doublet and triplets as in Eq.~(\ref{eq:phivev}), and then derive the complete set of extremum conditions, namely,
\begin{eqnarray}
		0 &=& \frac{\partial V^{}_{\rm soft}}{\partial  v{}_{\rm H}}= -\mu^{2}_{\rm H}v^{}_{\rm H} + \lambda^{}_{\rm H}v^3_{\rm H} + \lambda^{}_{14} \left(v^2_{1} + v^2_{2}\right) v^{}_{\rm H} - 2\sqrt{2}\mu^{}_1  v^{}_{\rm H} (v^{}_1 \cos\alpha^{}_1+ v^{}_2 \cos\alpha^{}_2) \; , \nonumber \\
		0 &=& \frac{\partial V^{}_{\rm soft}}{\partial  v{}_{1}} =  2m^{2}_{11} v^{}_1 +4\lambda^{}_1 \left(v^2_{1} + v^2_{2}\right) v^{}_1 + \lambda^{}_{14}v^2_{\rm H}v^{}_1 - \sqrt{2}\mu^{}_1 v^2_{\rm H}\cos\alpha^{}_1\; , \nonumber \\
		0 &=& \frac{\partial V^{}_{\rm soft}}{\partial  v{}_{2}} =  2m^{2}_{11} v^{}_2 +4\lambda^{}_1 \left(v^2_{1} + v^2_{2}\right) v^{}_2 + \lambda^{}_{14}v^2_{\rm H}v^{}_2 - \sqrt{2}\mu^{}_1 v^2_{\rm H}\cos\alpha^{}_2 \; , \nonumber \\
		0 &=& \frac{\partial V^{}_{\rm soft}}{\partial  \alpha^{}_1} = \sqrt{2} \mu^{}_1 v^2_{\rm H}v^{}_{1}\sin\alpha^{}_1  \; , \nonumber \\
		0 &=& \frac{\partial V^{}_{\rm soft}}{\partial  \alpha^{}_2} = \sqrt{2} \mu^{}_1 v^2_{\rm H}v^{}_{2}\sin\alpha^{}_2  \; .
		\label{eq:softvevcond}
\end{eqnarray}
The last two equalities in Eq.~(\ref{eq:softvevcond}) indicate $\alpha^{}_1,\alpha^{}_2  = 0$ or $\pi$, namely, there is no spontaneous CP violation in $V^{}_{\rm soft}$ when the Higgs fields acquire their individual vev's\cite{Ferreira:2021bdj}. Without loss of generality, we set $\alpha^{}_1= \alpha^{}_2 =  0$ in the following. Then the solutions to Eq.~(\ref{eq:softvevcond}) can be divided into two classes. If $v^{}_1 \neq v^{}_2$ holds, we obtain
\begin{eqnarray}
			v^{}_{\rm H} = 0 \; , \quad v^2_1 =- \frac{m^2_{11}}{2\lambda^{}_1}\cos^2_{}\theta  \; , \quad
			v^2_2 = - \frac{m^2_{11}}{2\lambda^{}_1}\sin^2_{}\theta \; ,
\label{eq:softvev1}
\end{eqnarray}
which is apparently not of our interest. If $v^{}_1 = v^{}_{2}$ is assumed, Eq.~(\ref{eq:softvevcond}) is reduced to
\begin{eqnarray}
			-\mu^{2}_{\rm H} + \lambda^{}_{\rm H} v^2_{\rm H} + 2\lambda^{}_{14} v^2_1 - 4\sqrt{2} \mu^{}_1  v^{}_1 &=& 0 \; , \nonumber \\
			\lambda^{}_{14}v^2_{\rm H} v^{}_1 - \sqrt{2} \mu^{}_1  v^2_{\rm H} + 2m^{2}_{11} v^{}_1 + 8\lambda^{}_1 v^3_1 &=& 0 \; .
\label{eq:softvevreduce}
\end{eqnarray}
		
As we have mentioned before, $\mu^{}_1$ should be a small parameter compared to the other mass parameters, rendering the perturbative solutions to Eq.~(\ref{eq:softvevreduce}) to be possible. Up to the first order of $\mu^{}_1 $, {\bf Type A} vacuum solution given in Eq.~(\ref{eq:so4vevsol1}) is modified to be
\begin{eqnarray}
			v^2_{\rm H,i} &=& \frac{4\mu^{2}_{\rm H}\lambda^{}_1 + 2m^2_{11} \lambda^{}_{14}}{4\lambda^{}_{\rm H}\lambda^{}_1-\lambda^2_{14}}-\frac{4\left(8m^{2}_{11}\lambda^{}_{\rm H}\lambda^{}_1 + m^{2}_{11} \lambda^2_{14}+6\mu^2_{\rm H}\lambda^{}_1 \lambda^{}_{14}\right)}{\sqrt{\left(2m^2_{11}\lambda^{}_{\rm H}+\mu^2_{\rm H}\lambda^{}_{14}\right)\left(\lambda^2_{14}-4\lambda^{}_{\rm H}\lambda^{}_1\right)^3_{}}}\mu^{}_1  \; , \nonumber \\
			v^{}_{1(2),{\rm i}}  &=& \sqrt{\frac{2m^{2}_{11}\lambda^{}_{\rm H}+\mu^{2}_{\rm H} \lambda^{}_{14}}{2\left(\lambda^2_{14}-4\lambda^{}_{\rm H}\lambda^{}_1\right)}}-\frac{\sqrt{2}\left(2\mu^2_{\rm H}\lambda^{}_{\rm H}\lambda^{}_1+\mu^2_{\rm H}\lambda^2_{14} + 3m^2_{11} \lambda^{}_{\rm H}\lambda^{}_{14}\right)}{\left(2m^2_{11} \lambda^{}_{\rm H}+\mu^2_{\rm H}\lambda^{}_{14}\right)\left(4\lambda^{}_{\rm H}\lambda^{}_{1}-\lambda^2_{14}\right)}\mu^{}_1  \; ,
\label{eq:softsol1}
\end{eqnarray}
or
\begin{eqnarray}
			v^2_{\rm H,ii} &=& \frac{4\mu^{2}_{\rm H}\lambda^{}_1 + 2m^2_{11} \lambda^{}_{14}}{4\lambda^{}_{\rm H}\lambda^{}_1-\lambda^2_{14}}+\frac{4\left(8m^{2}_{11}\lambda^{}_{\rm H}\lambda^{}_1 + m^{2}_{11} \lambda^2_{14}+6\mu^2_{\rm H}\lambda^{}_1 \lambda^{}_{14}\right)}{\sqrt{\left(2m^2_{11}\lambda^{}_{\rm H}+\mu^2_{\rm H}\lambda^{}_{14}\right)\left(\lambda^2_{14}-4\lambda^{}_{\rm H}\lambda^{}_1\right)^3_{}}}\mu^{}_1  \; , \nonumber \\
			v^{}_{1(2),{\rm ii}}  &=& -\sqrt{\frac{2m^{2}_{11}\lambda^{}_{\rm H}+\mu^{2}_{\rm H} \lambda^{}_{14}}{2\left(\lambda^2_{14}-4\lambda^{}_{\rm H}\lambda^{}_1\right)}} - \frac{\sqrt{2}\left(2\mu^2_{\rm H}\lambda^{}_{\rm H}\lambda^{}_1+\mu^2_{\rm H}\lambda^2_{14} + 3m^2_{11} \lambda^{}_{\rm H}\lambda^{}_{14}\right)}{\left(2m^2_{11} \lambda^{}_{\rm H}+\mu^2_{\rm H}\lambda^{}_{14}\right)\left(4\lambda^{}_{\rm H}\lambda^{}_{1}-\lambda^2_{14}\right)}\mu^{}_1  \; .
\end{eqnarray}
We can also calculate the difference between the minima of $V^{}_{\rm soft}$ corresponding to the above two vacuum solutions, namely,
\begin{eqnarray}
			\Delta V^{\rm min}_{\rm soft} = V^{\rm min}_{\rm soft,i} - V^{\rm min}_{\rm soft,ii} = \frac{8\left(2\mu^2_{\rm H}\lambda^{}_1 + m^2_{11} \lambda^{}_{14}\right)}{\lambda^2_{14}-4\lambda^{}_{\rm H}\lambda^{}_1}\sqrt{\frac{2m^2_{11}\lambda^2_{\rm H}+\lambda^{}_{14}\mu^2_{\rm H}}{\lambda^{2}_{14} - 4\lambda^{}_{\rm H}\lambda^{}_1}}\mu^{}_1  + {\cal O}(\mu^2_{1}) \; .
\label{eq:vsoftmin}
\end{eqnarray}
Hence whether $\{v^2_{\rm H,i}, v^{}_{1(2), {\rm i}}\}$ or $\{v^2_{\rm H,ii}, v^{}_{1(2), {\rm ii}}\}$ is the global minimum of $V^{}_{\rm soft}$ actually depends on the sign of $\mu^{}_1 $.

\begin{figure}[t!]

	\centering		\includegraphics[width=\textwidth]{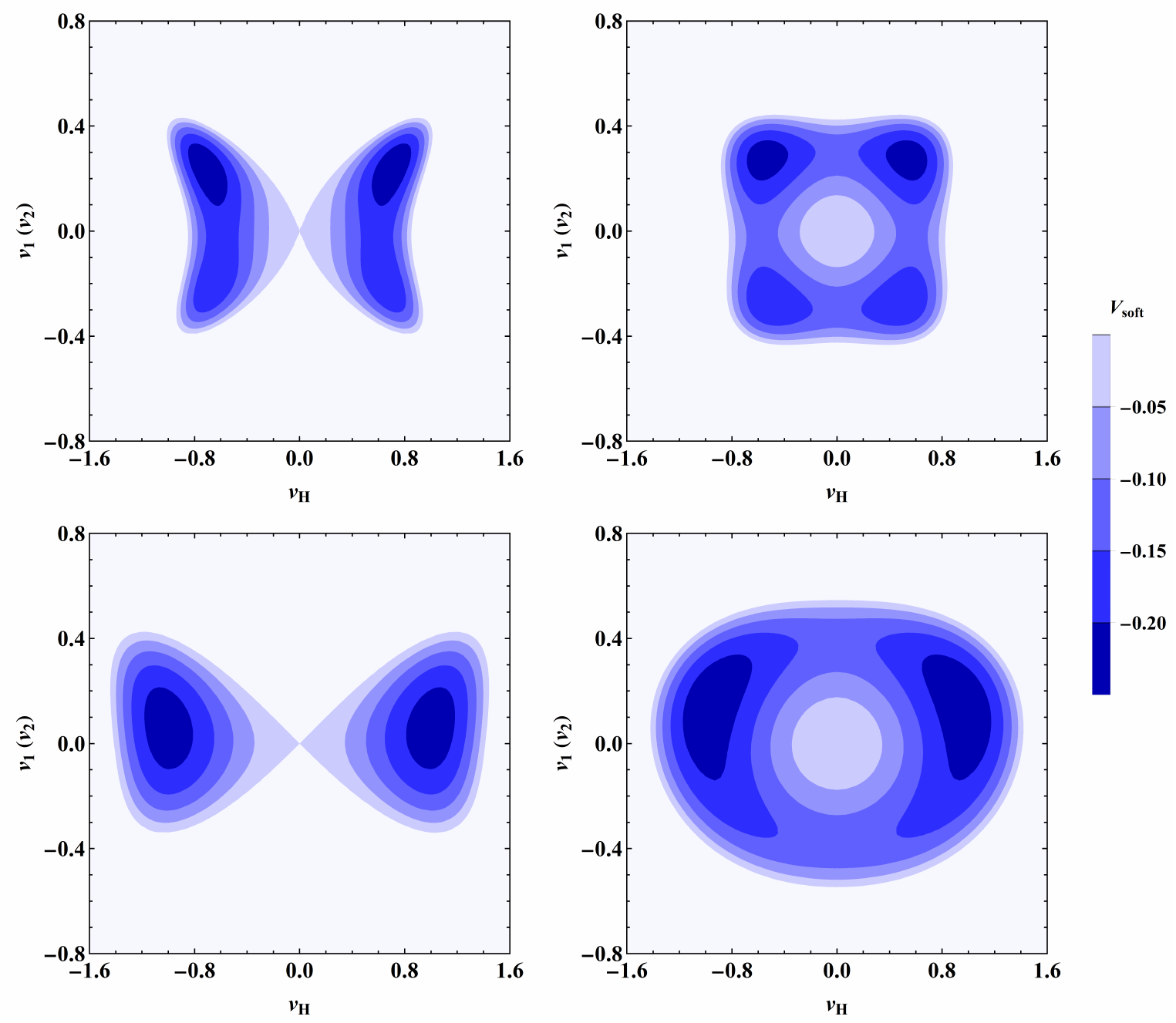} 	
	\caption{The distributions of $V^{}_{\rm soft}$ with respect to $v^{}_{\rm H}$, $v^{}_1$ and $v^{}_2$ in four different case with typical parameters. {\it Top-left panel}: $\mu^{2}_{\rm H}=1.99$, $m^{2}_{11}=0.33$, $\lambda^{}_{\rm H}=5.48$, $\lambda^{}_{1}=5.48$, $\lambda^{}_{14}=-6.03$, $\mu^{}_1 = 0.07$; {\it Top-right panel}: $\mu^{2}_{\rm H}=1.41$, $m^{2}_{11}=-1.49$, $\lambda^{}_{\rm H}=4.12$, $\lambda^{}_{1}=4.12$, $\lambda^{}_{14}=1.65$, $\mu^{}_1 = 0.07$; {\it Bottom-left panel}: $\mu^{2}_{\rm H}=0.90$, $m^{2}_{11}=0.90$, $\lambda^{}_{\rm H}=0.90$, $\lambda^{}_{1}=0.90$, $\lambda^{}_{14}=-0.68$, $\mu^{}_1 = 0.05$; {\it Bottom-right panel}: $\mu^{2}_{\rm H}=0.90$, $m^{2}_{11}=-0.90$, $\lambda^{}_{\rm H}=0.90$, $\lambda^{}_{1}=1.49$, $\lambda^{}_{14}=2.09$, $\mu^{}_1 = 0.05$. All the mass-dimensional parameters have been properly normalized to be dimensionless.}
	\label{fig:vsoft} 
	\vspace{0cm}
\end{figure}
		
On the other hand, the corrections from $\mu^{}_1$ to {\bf Type B} vacuum solution can be calculated in a similar way. The vev's of $H$ and ${\bm \phi}^{}_i$ up to the second order of $\mu^{}_1$ are found to be
\begin{eqnarray}
v^2_{\rm H, iii} &=& \frac{\mu^2_{\rm H}}{\lambda^{}_{\rm H}} \left[1 + \frac{4\left(\mu^2_{\rm H}\lambda^{}_{14}+4m^2_{11}\lambda^{}_{\rm H}\right)}{\left(\mu^2_{\rm H}\lambda^{}_{14}+2m^2_{11}\lambda^{}_{\rm H}\right)^2_{}} \mu^2_{1} \right] \; , \nonumber \\
			v^{}_{1(2),{\rm iii}} &=& \frac{\sqrt{2}\mu^2_{\rm H}\mu^{}_1 }{\mu^2_{\rm H} \lambda^{}_{14} + 2 m^2_{11} \lambda^{}_{\rm H}} \; ,
\label{eq:softsol3}
\end{eqnarray}
where $v^{}_{1}$ and $v^{}_{2}$ are proportional to the trilinear mass parameter $\mu^{}_1$, thus the small magnitude of $\mu^{}_1$ naturally induces small vev's of the Higgs triplets. Furthermore, if the coefficients of the quartic terms involving Higgs triplets in $V^{}_{\rm soft}$ are all vanishing, then Eq.~(\ref{eq:softsol3}) will be reduced to
\begin{eqnarray}
v^2_{\rm H,iii} = \frac{\mu^2_{\rm H
		}}{\lambda^{}_{\rm H}}\left(1+\frac{4\mu^{2}_1 }{m^2_{11} \lambda^{}_{\rm H}}\right) \; , \quad
		v^{}_{\rm 1(2),iii} =  \frac{\mu^2_{\rm H}\mu^{}_1 }{\sqrt{2} m^2_{11} \lambda^{}_{\rm H}} \; ,
		\label{eq:softsol3red}
\end{eqnarray}
which are in accord with the familiar formulas of the vev's of $H$ and ${\bm \phi}^{}_i$ in the conventional type-II seesaw model.
	
In Fig.~\ref{fig:vsoft}, we show the distributions of the scalar potential $V^{}_{\rm soft}$ with respect to $v^{}_{\rm H}$, $v^{}_1$ and $v^{}_2$, where the relevant parameters are consistent with those in Fig.~\ref{fig:vso4} except for nonzero values of $\mu^{}_1$. Comparing the results in Fig.~\ref{fig:vso4} and Fig.~\ref{fig:vsoft}, we can clearly see that the modifications from soft symmetry-breaking terms slightly reshape the potential. In particular, the degeneracy of the vacua is partially broken.
	
The inclusion of two extra Higgs triplets in the 2HTM extends also the scalar mass spectrum. As for the scalar potential $V^{}_{\rm soft}$, after the spontaneous symmetry breaking, we have three CP-even and two CP-odd neutral scalars, whose masses are given by
\begin{eqnarray}
		m^2_{h^0_{}} & \approx & 2v^2_{\rm H}\left(\lambda^{}_{\rm H} +  \frac{4\lambda^{}_1 v^2_1}{v^2_{\rm H}}\right) \; , \nonumber \\
		m^2_{H^0_1} &\approx& m^2_{H^0_2} = \frac{\sqrt{2}v^2_{\rm H} \mu^{}_1}{v^{}_1} \; , \nonumber \\
		m^2_{A^0_{1}} &= & \frac{\sqrt{2}v^2_{\rm H} \mu^{}_1}{v^{}_1} \; , \nonumber \\
		m^2_{A^0_{2}} &=&  \frac{\sqrt{2}v^2_{\rm H} \mu^{}_1}{v^{}_1}\left(1 + \frac{8 v^2_1}{v^2_{\rm H}} \right) \; ,
\label{eq:neumass}
\end{eqnarray}
where $m^{}_{h^0_{}}$ turns out to be the SM-like Higgs boson of mass $m^{}_{h^0_{}} \approx 125~{\rm GeV}$. Meanwhile, there are two singly-charged and two doubly-charged scalars. To the second order of $v^{}_{1}/v^{}_{\rm H}$, we have
\begin{eqnarray}
    	m^2_{H^\pm_1} &=& \frac{\sqrt{2}v^2_{\rm H}\mu^{}_1}{v^{}_1}\left(1 + \frac{8 v^2_1}{v^2_{\rm H}}\right) \; , \nonumber \\
    	m^2_{H^\pm_2} &\approx&  m^2_{H^{\pm\pm}_1} \approx m^2_{H^{\pm\pm}_2} = \frac{\sqrt{2}v^2_{\rm H} \mu^{}_1}{v^{}_1} \; .
\label{eq:chargemass}
\end{eqnarray}
From Eqs.~(\ref{eq:neumass}) and (\ref{eq:chargemass}), we can see that the masses of $H^0_{1,2}$, $H^\pm_{1,2}$, $H^{\pm\pm}_{1,2}$ and $A^0_{1,2}$ are nearly degenerate. In addition, the doubly-charged Higgs bosons have been restricted to be heavier than $800~{\rm GeV}$ by the direct searches for same-sign dileptons at the LHC~\cite{ATLAS:2017xqs,CERNReport}. This constraint can be easily satisfied by adjusting the values of $\mu^{}_1$ and $v^{}_1$ in our model.

In our calculations, we have assumed the vev's of all the Higgs fields to be neutral {\it a priori}. However, the charge-breaking and neutral vacua may coexist in the Higgs-triplet models~\cite{Arhrib:2011uy, Arhrib:2011vc, Xu:2016klg, BhupalDev:2018xya, Ferreira:2019hfk}. Therefore, it may be necessary to examine all the possible vacua and make sure that the neutral one is indeed the deepest. In addition, the spontaneous breaking of accidental symmetries in the early Universe may result in some interesting topological defects of the vacuum structure, such as domain walls, vortices and monopoles~\cite{Achucarro:1999it,Battye:2011jj,Vilenkin:1994,Eto:2018hhg,Eto:2020hjb,Law:2021ing,Eto:2019hhf}. We shall leave these relevant issues in the 2HTM for future works.

\section{Summary}\label{sec:sum}

As is well known, tiny Majorana neutrino masses can be explained in the 2HTM via the type-II seesaw mechanism, while the cosmological matter-antimatter asymmetry can be generated in the same framework by the triplet leptogenesis. In this paper, we explore all possible accidental symmetries in the scalar potential of the 2HTM. Our main results can be summarized as follows.
	
First, we write down the most general Lagrangian in the 2HTM. Adopting the bilinear-field formalism, we construct a ten-dimensional bilinear vector $R^\mu_{}$ and rewrite the pure-triplet potential in the quadratic form of $R^\mu_{}$. Based on the group-theoretical arguments, we demonstrate that the maximal symmetry group that the 2HTM potential can accommodate is ${\rm SO(4)}$. The nine spatial components of $R^\mu_{}$ then form a nine-dimensional representation of ${\rm SO} (4)$. By carefully analyzing the subgroups of ${\rm SO} (4)$, we arrive at totally eight kinds of accidental symmetries. As the doublet-triplet-mixing terms are vitally important in a realistic model, we also discuss their influence on the accidental symmetries. It is found that the existence of trilinear terms $H^{\rm T}_{}{\rm i}\sigma^{}_2{\bm \sigma}\cdot {\bm \phi}^{}_{i}  H$ severely violates the accidental symmetries in the 2HTM. For each of these symmetries, we derive explicit relations among the mass parameters and coupling constants in the scalar potential, which are listed in Table~\ref{Table:classification}.
	
Second, the scalar potential should be bounded from below to ensure the existence and stability of vacua. For the pure-triplet potential, sufficient BFB conditions are derived by requiring the coupling coefficient matrix $L^{}_{\mu\nu}$ to be positive definite. We show that such BFB conditions can be simplified if the potential maintains some specific accidental symmetries. In particular, we obtain the sufficient and necessary conditions for the complete ${\rm SO(4)}$-invariant potential to be bounded from below.
	
Finally, as an application of the accidental symmetries, we discuss the vacuum structure of the ${\rm SO(4)}$-invariant potential. In the absence of soft symmetry-breaking terms, there are two types of neutral vacuum solutions with $v^{}_{\rm H} \neq 0$. For {\bf Type A} vacuum solution, both $v^{}_{\rm H}$ and $v^{}_{1(2)}$ are non-zero, and an ${\rm SO(2)}$ residual symmetry exists in the flavor space of two Higgs triplets. We explain why {\bf Type A} solution is unlikely to be the realistic vacuum although it may give correct magnitudes of $v^{}_{\rm H}$, $v^{}_{1}$ and $v^{}_{2}$. On the other hand, {\bf Type B} vacuum solution corresponds to the vanishing vev's of Higgs triplets, which however will receive corrections from the soft symmetry-breaking terms and become non-zero. We show how the soft symmetry-breaking terms modify the vacuum solutions. Furthermore, the scalar mass spectrum has been figured out.
	
Accidental symmetries in the 2HTM potential will be useful to reduce the number of free model parameters, and thus enhance the predictive power of the theory. In this connection, we believe that the systematic classification of accidental symmetries in the 2HTM potential in the present work will be valuable for future phenomenological studies.

\section*{Acknowledgements}
The authors thank Xun-jie Xu, Bingrong Yu and Di Zhang for helpful discussions. This work was supported in part by the National Natural Science Foundation of China under grant No. 11775232 and No. 11835013, by the Key Research Program of the Chinese Academy of Sciences under grant No. XDPB15, and by the CAS Center for Excellence in Particle Physics.

\newpage
\begin{landscape}	
\begin{table}
\begin{small}
\centering
\hspace{0.4cm}
\begin{tabular}{c|c|ccccccccc}
				\toprule
				\hline
				 Symmetry & Generators & $\makecell[c]{m_{22}^2}$ & $\makecell[c]{m_{12}^2}$ & $\makecell[c]{\lambda_2^{}}$  & $\makecell[c]{\lambda_3^{}}$ & $\makecell[c]{\lambda_4^{}}$ & $\makecell[c]{{\rm Re}\, \lambda_5^{} }$ & $\lambda_6^{} = \lambda_7^{}$  & $\lambda_{10}^{}$ & ${\rm Re}\, \lambda_{11}^{}$\\
				\hline
${\rm SO(4)}$ & $J^{1,2,3,4,5,6}_{}$ & $\makecell[c]{m_{11}^2}$  & 0 & $\lambda_1^{}$  & $2\lambda_1^{} - 2\lambda_8^{}$ & $2\lambda_8^{}$ & $\makecell[c]{0}$ & 0 &  $2\lambda_8^{}$ & 0 \\
				\hline
${\rm O(3)}^i_{} \otimes {\rm O(2)}^j_{}$ & $J^{1,2,3,4}_{}$ & $\makecell[c]{m_{11}^2}$  & 0 & $\lambda_1^{}$  & $2\lambda_1^{} - \lambda_4^{}$ & -- & $\makecell[c]{0}$ & 0 &  $2\lambda_8^{}$ & 0 \\
				\hline
				 &$J^{1,4,5,6}_{}$ & $\makecell[c]{m_{11}^2}$   & 0 & $\lambda_1^{}$  & -- & $2\lambda_8^{}$ & $\makecell[c]{0}$ & 0 &  $2\lambda_1^{} - \lambda_3^{}$ & 0 \\
$\makecell[c]{ {\rm O(2)}^i_{} \otimes {\rm O(3)}^j_{}}$ & $J^{2,4,5,6}_{}$ &  $\makecell[c]{m_{11}^2}$ & 0 & $\lambda_1^{}$  & -- & $2\lambda_8^{}$ & $-2\lambda_1^{} + \lambda_3^{} + 2\lambda_8^{}$ & 0 &  $2\lambda_1^{} - \lambda_3^{}$ & $ \displaystyle +\lambda_1^{} - \lambda_8^{} -\lambda_3^{}/2$ \\
				& $J^{3,4,5,6}_{}$ & $\makecell[c]{m_{11}^2}$  & 0 & $\lambda_1^{}$  & -- & $2\lambda_8^{}$ & $+2\lambda_1^{} - \lambda_3^{}- 2\lambda_8^{} $ & 0 &  $2\lambda_1^{} - \lambda_3^{}$ & $ \displaystyle - \lambda_1^{} + \lambda_8^{} +\lambda_3^{}/2$ \\
				\hline
${\rm O(3)}^j_{} \otimes Z_2^{}$ & $J^{4,5,6}_{}$ & $\makecell[c]{m_{11}^2}$  & 0 & $\lambda_1^{}$  & -- & $2\lambda_8^{}$ & $-2\,{\rm Re}\, \lambda_{11}^{}$ & 0 &  $2\lambda_1^{} - \lambda_3^{}$ & -- \\
				\hline
				& $J^{1,4}_{}$  & -- & 0 & -- & -- & -- & 0 & 0 & -- & 0 \\
$\makecell[c]{ {\rm O(2)}^i_{} \otimes {\rm O(2)}^j_{}}$ & $J^{2,4}_{}$ &  $\makecell[c]{m_{11}^2}$  & ${\rm Im}\, m_{12}^2$ & $\lambda_1^{}$  & -- & -- & $+2\lambda_1^{}  - \lambda_3^{} - \lambda_4^{}$ & ${\rm Im}\, \lambda_6^{}$ &  $2\lambda_8^{} - 2\,{\rm Re}\, \lambda_{11}^{}$ & -- \\
				& $J^{3,4}_{}$ & $\makecell[c]{m_{11}^2}$  & ${\rm Re}\, m_{12}^2$ & $\lambda_1^{}$  & -- & -- & $- 2\lambda_1^{} + \lambda_3^{} + \lambda_4^{} $ & ${\rm Re}\, \lambda_6^{}$ &  $2\lambda_8^{} + 2\,{\rm Re}\, \lambda_{11}^{}$ & -- \\
				\hline
				& $J^{1,4}_{}$ &$\makecell[c]{m_{11}^2}$ & 0 & $\lambda_1^{}$  & -- & -- & 0 & 0 &  -- & 0 \\
$\makecell[c]{ {\rm O(2)}^i_{} \otimes {\rm O(2)}^j_{} \otimes Z_2^{}}$ & $J^{2,4}_{}$ & $\makecell[c]{m_{11}^2}$  & 0 & $\lambda_1^{}$  & -- & -- & $ +2\lambda_1^{} - \lambda_3^{} - \lambda_4^{}$ & 0 &  $2\lambda_8^{} - 2\,{\rm Re}\, \lambda_{11}^{}$ & -- \\
				& $J^{3,4}_{}$ & $\makecell[c]{m_{11}^2}$  & 0 & $\lambda_1^{}$  & -- & -- & $- 2\lambda_1^{}+ \lambda_3^{} + \lambda_4^{} $ & 0 &  $2\lambda_8^{} + 2\,{\rm Re}\, \lambda_{11}^{}$ & -- \\
				\hline
${\rm SO(2)}^j_{} \otimes (Z_2^{})^2_{}$ & $J^{4}_{}$ & $\makecell[c]{m_{11}^2}$  & 0 & $\lambda_1^{}$  & -- & -- & -- & 0 & -- &  -- \\
				\hline
				&  & $\makecell[c]{m_{11}^2}$   & -- & -- & -- & -- & -- & -- & -- &  -- \\
${\rm O(2)}^j_{} \otimes Z_2^{}$ & $J^{4}_{}$   & -- & -- & $\lambda_1^{}$  & -- & -- & -- & -- & -- &  -- \\
				&   & -- & 0  & -- & -- & -- & -- & 0 & -- &  -- \\
				\hline
				\bottomrule
			\end{tabular}
		\end{small}	
\vspace{0.5cm}			
\caption{Eight types of accidental symmetries that can exist in the 2HTM potential, together with the corresponding generators and restrictions on the relevant parameters. A dash indicates that there is no constraint on the parameter. Apart from the relations shown in this table, ${\rm Im}\, \lambda_5^{} = 0$, $\lambda_8^{} = \lambda_9^{}$, ${\rm Im}\, \lambda_{11}^{} = 0$ and $\lambda_{12}^{} = \lambda_{13}^{} = 0$ should also be satisfied for all of these symmetries. Constraints on the parameters in $V^{}_{{\rm H}\phi}$ are not presented here, for which one can refer to Sec.~\ref{subsec:ent}.}
			\label{Table:classification}
		\end{table}
	\end{landscape}

\end{document}